\documentclass{openjournal}
\usepackage{natbib}
\usepackage{bm}

\renewcommand{\it}[1]{\textit{#1}}
\usepackage[dvipsnames]{xcolor}
\usepackage{graphicx}	
\usepackage{amsmath}	
\usepackage[linktocpage=true]{hyperref}
\usepackage{setspace}
\usepackage{verbatim}
\usepackage{orcidlink}
\usepackage{makecell}
\usepackage{changepage}
\usepackage[toc,page]{appendix}
\usepackage{lineno}

\renewcommand{\arraystretch}{2}

\usepackage[xindy]{glossaries-extra}
\makeglossaries
\newglossaryentry{ellipticity}
{
    name=ellipticity,
    description={A quantity describing the axis ratio and orientation of a 2D ellipse.}
}

\newglossaryentry{tidal field}
{
    name=tidal field,
    description={The gradient of gravitational forces created by large-scale structure.}
}

\newglossaryentry{estimator}
{
    name=estimator,
    description={An estimator for a quantity is a function of the data. When evaluated at the observed data it provides an estimate of the true value of the quantity.}
}

\newglossaryentry{halo}
{
    name=halo,
    description={Usually refers to the dark matter ``clump" surrounding a galaxy or cluster of galaxies.}
}

\newglossaryentry{tracer}
{
    name=tracer,
    description={An observable that can be used to trace, or map, the underlying dark matter distribution. For IA this is often the position of galaxies. All tracers are imperfect and have some bias that needs to be accounted for.}
}

\newglossaryentry{radial}
{
    name=radial,
    description={The direction along the \gls{LOS}, or $\Pi$. As opposed to ``transverse", or $r_p$.}
}

\newglossaryentry{transverse}
{
    name=transverse,
    description={The direction perpendicular to the line of sight, or $r_p$.}
}

\newglossaryentry{shear}
{
    name=shear,
    description={The anisotropic stretch of a galaxy in a given direction.}
}

\newglossaryentry{cosmic shear}
{
    name=cosmic shear,
    description={The coherent distortion of galaxy images through weak gravitational lensing by the large-scale  structure along the LOS.}
}

\newglossaryentry{weak lensing}
{
    name=weak lensing,
    description={Gravitational lensing in the regime where image distortions are small and appear only in the statistical ensemble of galaxies.}
}

\newglossaryentry{tidal alignment}
{
    name=tidal alignment,
    description={Refers to how intrinsic galaxy ellipticities are correlated with the underlying tidal field.}
}

\newglossaryentry{correlation function}
{
    name=correlation function,
    description={A statistical quantity that measures the degree of correlation between two random variables. In weak lensing, it quantifies the correlation between galaxy shapes as a function of their separation.}
}

\newglossaryentry{fundamental plane}
{
    name=fundamental plane,
    description={This represents the global properties of galaxies, such as luminosity, radius, projected velocity dispersion, and projected luminosity. It is defined in terms of their velocity dispersion and mean surface brightness.}
}

\newglossaryentry{ELG}
{
    name=ELG,
    description={Emission-Line Galaxies are a type of galaxy that emit strong light at specific wavelengths, typically in the optical and near-infrared parts of the electromagnetic spectrum.}
}

\newglossaryentry{LRG}
{
    name=LRG,
    description={Luminous Red Galaxies are characterized by high luminosity in the red part of the electromagnetic spectrum. Comparable to elliptical galaxies.}
}

\newglossaryentry{satellite}
{
    name=satellite,
    description={Small galaxies in a halo that orbit larger, more massive central galaxies.}
}

\newglossaryentry{central}
{
    name=central,
    description={The central, typically largest and most massive, galaxy in a cluster or halo.}
}

\newglossaryentry{late-type galaxies}
{
    name=late-type galaxies,
    description={These are usually spiral galaxies, as described by the Hubble Sequence \citep{hubble_extragalactic_1926} and are sometimes referred to as simply ``blue" galaxies.}
}

\newglossaryentry{early-type galaxies}
{
    name=early-type galaxies,
    description={These are usually elliptical galaxies, as described by the Hubble Sequence \citep{hubble_extragalactic_1926}, and are sometimes referred to as simply ``red" galaxies.}
}

\newglossaryentry{$E$-mode}
{
    name= $E$-mode,
    description={Curl-free component of the shear field, in analogy to the electric field. The weak lensing shear field should contain only $E$-modes.}
}

\newglossaryentry{$B$-mode}
{
    name= $B$-mode,
    description ={Gradient-free component of the shear field, in analogy to the magnetic field. The detection of $B$-modes in weak lensing data can indicate the presence of systematic effects such as IA.}
}

\newglossaryentry{Bessel function}
{
    name = Bessel function,
    description=
    {Bessel functions are generalizations of trigonometric functions. Bessel functions of the first kind $J_\nu$ are solutions of the second-order differential equation
    \begin{equation*}
        x^2y^{\prime\prime} +xy^\prime +(x^2-\nu^2)y=0\,,
    \end{equation*}
    which are well-behaved at $x=0$. $\nu$ is a constant.     
    Spherical Bessel functions $j_\nu$ are solutions of
    \begin{equation*}
        x^2y^{\prime\prime}+2xy^\prime +(x^2-\nu(\nu+1))y=0\,.\notag
    \end{equation*}}
}

\newglossaryentry{LOS}
{
    name=LOS,
    description={``Line of Sight", i.e. radial direction away from Earth. For a helpful guide to distance measurements in cosmology, see~\cite{hogg_distance_2000}.}
}

\newglossaryentry{RSD}
{
    name=RSD,
    description={Redshift-Space Distortions describe the anisotropic clustering that is present when mapping out the redshift-space positions of galaxies, as opposed to their true positions. It is caused by the peculiar velocities of galaxies within clusters and, on larger scales, the growth of structure.}
}

\newglossaryentry{hydrodynamic simulations}
{
    name=hydrodynamic simulations,
    description={Simulations of galactic and structure formation that contain baryons and feedback. Done on smaller scales than N-body simulations which are dark matter only.}
}

\newglossaryentry{elliptical}
{
    name=elliptical,
    description={Large galaxies with little gas and dust. Similar to the ``early-type" and red galaxies.}
}

\newglossaryentry{spiral}
{
    name=spiral,
    description={Galaxies with distinct central bulges and spiral arms. Similar to the ``late-type" and blue galaxies.}
}

\newglossaryentry{torque}
{
    name=torque,
    description={A force vector $\mathbf{F}$ induces torque $\mathbf{\tau}$ with respect to a point of reference $A$ on an object $O$ such that
    \begin{equation}
        \boldsymbol{\tau} = \mathbf{r} \times \mathbf{F}\,,
    \end{equation}
    $\mathbf{r}$ being the vector connecting $A$ with $O$.}
}

\newglossaryentry{redshift}
{
    name=redshift,
    description={Cosmological redshift, $z$, refers to the fractional change of the wavelength of light, $\lambda_e$, emitted by a distant source and its observed wavelength when it reaches us, $\lambda_0$,
    \begin{equation*}
        z = \frac{\lambda_0 -\lambda_e}{\lambda_e}\,.
    \end{equation*}
This redshift is caused by the expansion of the universe.}
}

\newglossaryentry{photometric}
{
    name=photometric,
    description={In the context of cosmology, this refers to galaxy properties (most commonly redshifts) that were estimated through imaging color, as opposed to spectroscopy.}
}

\newglossaryentry{spectroscopic}
{
    name=spectroscopic,
    description={A spectroscopic survey involves obtaining detailed spectra of different objects at various wavelengths, which allows for obtaining accurate redshifts.}
}

\newglossaryentry{systematic errors}
{
    name=systematic errors,
    description={Errors that systematically bias observations or scientific conclusions derived from observations. Systematic errors affect the accuracy, not precision, of a measurement.}
}

\newglossaryentry{Legendre polynomial}
{
    name=Legendre polynomial,
    description={Legendre polynomials are solutions of the second-order differential equation
    \begin{equation*}
        (1-x^2)y^{\prime\prime}-2xy^\prime +n(n+1)y=0\,.
    \end{equation*}
    They arise as solutions of Laplace's equation $\nabla f^2=0 $ in spherical coordinates.}
}

\newglossaryentry{protocloud}
{
    name=protocloud,
    description={Gas cloud that, following gravitational collapse, becomes a galaxy.}
}

\newglossaryentry{vorticity}
{
    name=vorticity,
    description={The curl of a vector field, $\mathbf{x}$, defined as
    \begin{equation}
       \mathbf{\omega} = \mathbf{\nabla}\times\mathbf{x}\,.
    \end{equation}
    Vorticity reflects the rotational state of a fluid element}
}

\newglossaryentry{trace-free}
{
    name=trace-free,
    description={A rank-2 tensor, $T_{ij}$, is said to be trace-free if
    \begin{equation}
        \sum_{i} T_{ii} = 0\,.
    \end{equation}
    The above is a useful property that can simplify mathematical operations and in certain cases help distinguish between physical processes encapsulated in the same tensor.}
}

\newglossaryentry{randoms}
{
    name=randoms,
    description={A ``randoms" catalog contains mock objects without a clustering signal, yet following the configuration of a survey in which the actual clustered tracers were observed.}
}

\newglossaryentry{Levi-Civita symbol}
{
name=Levi-Civita symbol,
description ={$\epsilon_{ijk}$ is defined as:
\begin{equation}
    \epsilon_{ijk}=\begin{cases} 
    +1 & \text{if  $(i,j,k)$ is an even permutation of (1,2,3)}\\
    -1 &\text{if  $(i,j,k)$ is an odd permutation of (1,2,3)}\quad \quad\,.\\
     \phantom{-}0 &\text{if any index is repeated}
    \end{cases}
\end{equation}}
}

\newglossaryentry{convergence}{
name = convergence,
description ={The isotropic increase or decrease in the observed size of a source image.}
}

\newglossaryentry{forward modeling}{
name = forward modeling,
description = {Forward modeling is a technique in which the expected distribution of data is inferred from parameters of a model that is assumed to be true (rather than the other way around).
An instructive discussion on forward modeling can be found in \cite{heavens_statistical_2009}.}
}

\newglossaryentry{pivot luminosity}{
name = pivot luminosity,
description = {The value of luminosity at which the luminosity function goes from one regime to another. This is set to correspond to the absolute magnitude of -22.}
}

\newglossaryentry{direct}{
name = direct,
description = {As in \textit{direct IA detection}. Estimation method of the intrinsic alignment amplitude only, from galaxy shapes.}
}

\newglossaryentry{indirect}{
name = indirect,
description = {As in \textit{indirect IA detection}. Estimation method of the intrinsic alignment amplitude, jointly with the cosmic shear signal.}
}

\newglossaryentry{mass-sheet degeneracy}{
name = mass-sheet degeneracy,
description = {A transformation of the surface mass density $\kappa$,
\begin{equation}
    \kappa \rightarrow \lambda \kappa + (1 - \kappa)\,,
\end{equation}
yields the same observed quantities, such as image shapes and positions, magnification ratios, luminosities, etc. This introduces what is called a ``mass-sheet degeneracy" since the value of $\kappa$ cannot be directly inferred.}
}

\newglossaryentry{thin-lens approximation}{
name = thin-lens approximation,
description = {A simplification of a lens system that assumes a negligible lens thickness so that a light ray is bent only once, on the lens plane.}
}

\newglossaryentry{galaxy-galaxy lensing}{
name = galaxy-galaxy lensing,
description = {The distortion of shapes of background (source) galaxies by individual foreground (lens) galaxies.}
}

\newglossaryentry{galaxy bias}{
name = galaxy bias,
description = {The difference in spatial clustering between galaxies and the underlying bulk matter distribution.}
}

\newglossaryentry{halo occupation distribution}{
name = halo occupation distribution,
description = {The bias between galaxy and halo mass. It is expressed in terms of the probability distribution that a halo of mass $M$ contains $N$ galaxies of a given type, together with information about the positions and velocities of galaxies and dark matter within the halo.}
}

\newglossaryentry{halo mass function}{
name = halo mass function,
description = {The number density of halos of a given mass.}
}
\GlsXtrEnablePreLocationTag{Page }{Pages }

\newcommand{\rfig}[1]{Figure~\ref{fig:#1}}

\newcommand{\req}[1]{Eq.~(\ref{eq:#1})}
\newcommand{\rsec}[1]{Section~\ref{sec:#1}}
\newcommand{\rssec}[1]{Section~\ref{subsec:#1}}

\newcommand{\headings}[1]{\vspace{.2in}
\noindent \textbf{#1}}

\definecolor{vlgray}{RGB}{245,245,245}
\definecolor{primary}{RGB}{255,140,0}
\definecolor{bg}{RGB}{250,250,250}
\definecolor{lgreen}{RGB}{184, 209, 126}
\definecolor{dgreen}{RGB}{112, 163, 0}
\definecolor{lorange}{RGB}{255, 199, 128}
\definecolor{dorange}{RGB}{255, 145, 0}
\definecolor{lblue}{RGB}{111, 181, 189}
\definecolor{dblue}{RGB}{0, 95, 117}

\hypersetup{
    colorlinks=true,
    linkcolor=dblue,
    citecolor=NavyBlue,
    urlcolor=lblue,
    pdftitle={IA-Guide},
    pdfpagemode=FullScreen,
    }

\def\bx{\bm{x}}
\def\bk{\bm{k}}
\renewcommand{\d}[1]{\ensuremath{\operatorname{d}\!{#1}}}

\shorttitle{The IA Guide}
\shortauthors{Lamman et al.}
\begin{document}
\title{The IA Guide:\\
A Breakdown of Intrinsic Alignment Formalisms}

\author{\vspace{.3cm}Claire Lamman\,\orcidlink{0000-0002-6731-9329}$^{1}$}
\author{Eleni Tsaprazi\,\orcidlink{0000-0001-5082-4380}$^{2,8}$}
\author{Jingjing Shi\,\orcidlink{0000-0001-9879-4926}$^{3}$}
\author{Nikolina Niko \v{S}ar\v{c}evi\'{c},\orcidlink{0000-0001-7301-6415}$^{4}$}
\author{Susan Pyne\,\orcidlink{0009-0008-9932-6241}$^5$}
\author{Elisa Legnani\,\orcidlink{0000-0001-7079-3796}$^{6}$}
\author{Tassia Ferreira\,\orcidlink{0000-0003-4016-3763}$^{7}$}

\affiliation{$^1$Center for Astrophysics $|$ Harvard \& Smithsonian, 60 Garden Street, Cambridge, MA 02138, USA}
\affiliation{$^2$The Oskar Klein Centre, Department of Physics, Stockholm University, Albanova University Center, SE 106 91 Stockholm, Sweden}
\affiliation{$^3$Kavli Institute for the Physics and Mathematics of the Universe (WPI), The University of Tokyo Institutes for Advanced Study (UTIAS), The University of Tokyo, 5-1-5 Kashiwanoha, Kashiwa-shi, Chiba, 277-8583, Japan}
\affiliation{$^4$School of Mathematics, Statistics and Physics, Newcastle University, Newcastle upon Tyne, NE1 7RU, UK}
\affiliation{$^5$Department of Physics and Astronomy, University College London, Gower Street, London WC1E 6BT, UK}
\affiliation{$^6$Institut de Física d’Altes Energies (IFAE), The Barcelona Institute of Science and Technology, Campus UAB, 08193 Bellaterra Barcelona, Spain}
\affiliation{$^7$Department of Physics, University of Oxford, Denys Wilkinson Building, Keble Road, Oxford OX1 3RH, UK}
\affiliation{$^8$Imperial Centre for Inference and Cosmology (ICIC) \& Astrophysics group, Department of Physics, Imperial College, Blackett Laboratory,
Prince Consort Road, London SW7 2AZ, UK\vspace{.5cm}}

\thanks{Corresponding Author: \href{mailto:claire.lamman@cfa.harvard.edu}{claire.lamman@cfa.harvard.edu}\\ \textit{For more details see \nameref{sec:contributions}}.}

\begin{abstract}
We summarize common notations and concepts in the field of Intrinsic Alignments (IA). IA refers to physical correlations involving galaxy shapes, galaxy spins, and the underlying cosmic web. Its characterization is an important aspect of modern cosmology, particularly in weak lensing analyses. This resource is both a reference for those already familiar with IA and designed to introduce someone to the field by drawing from various studies and presenting a collection of IA formalisms, estimators, modeling approaches, alternative notations, and useful references. 

\end{abstract}

\maketitle

\begin{figure}[h]
\centering
\vspace{.5cm}
\includegraphics[width=0.3\textwidth]{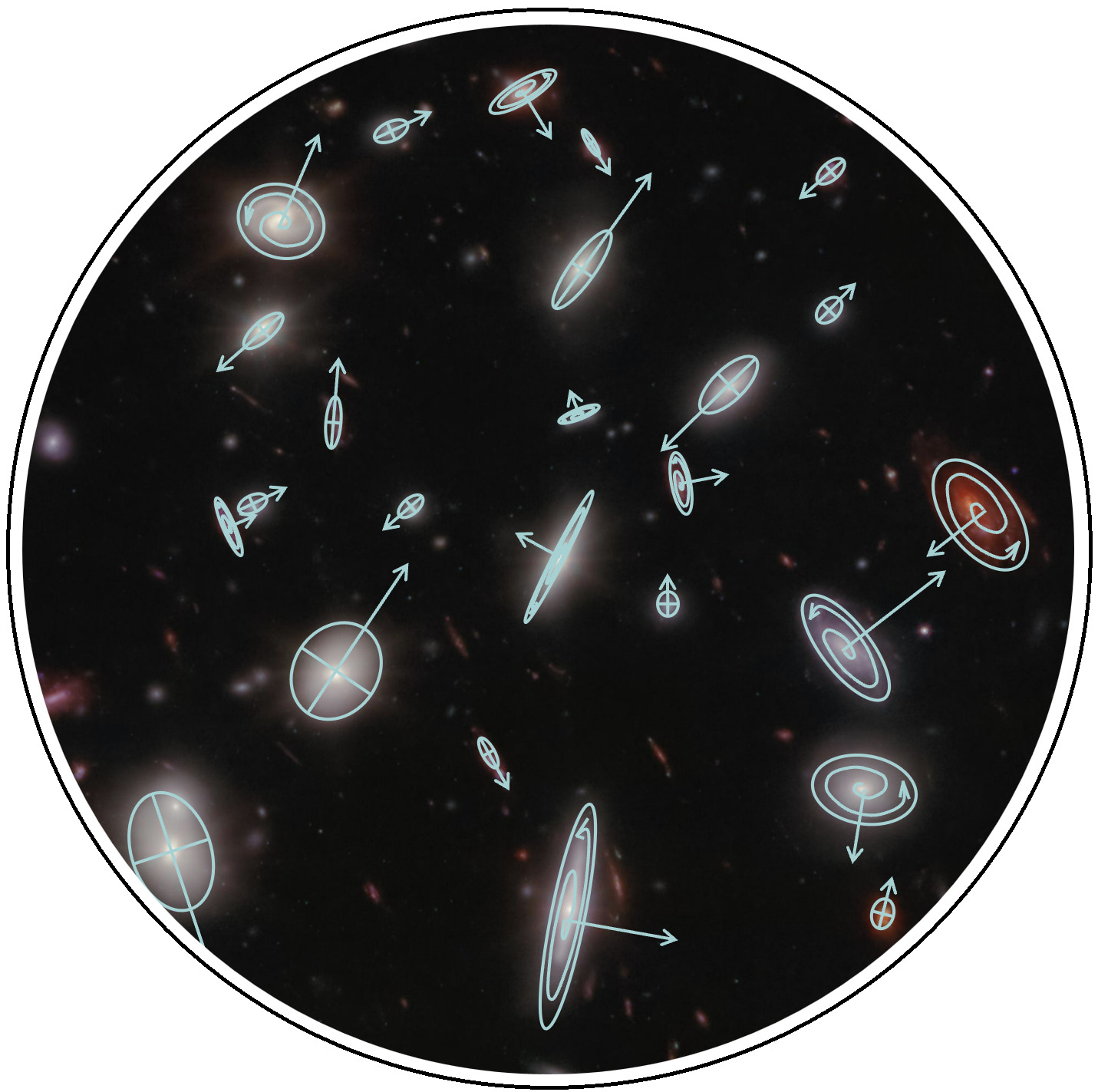}
\caption{\centering
\textit{Galaxy shapes and orientations traced over a portion of JWST's NIRCam image of Abell 2744.}
}
\label{fig:ia_doodle}
\end{figure}

\newpage

\begin{spacing}{1.3}
\tableofcontents
\end{spacing}
\addtocontents{toc}{}

\newpage
\section{Introduction}\label{sec:Introduction}
This resource is a condensed overview of quantities relevant for describing the \textit{intrinsic alignment} (IA) of galaxies.
For scientists new to the field, it is a useful starting place that contains a broad introduction to IA and helpful references with more details and derivations.
It is also structured to be a quick reference for those already familiar with IA.
This is not a review article and not necessarily intended to be read beginning-to-end.
Sections~\ref{sec:Ellipticity}-\ref{sec:shear_power_spectrum} each contain common formalisms of an IA \gls{estimator}, brief pedagogical explanations with practical advice, alternative notations, and useful references.
Sections \ref{sec:Modeling} and \ref{sec:ia_applications} summarize IA modeling and applications. Terms in \href{sec:glossary}{\textcolor{dblue}{teal}} are hyperlinked to glossary entries at the end of the document.

IA refers to correlations between galaxy shapes and between galaxy shapes and the underlying dark matter distribution (qualitatively illustrated in Figure \ref{fig:ia_doodle}). These arise naturally within our current understanding of galaxy formation, as confirmed by \gls{hydrodynamic simulations}~\citep{kiessling_galaxy_2015, bhowmick_evolution_2020, samuroff_advances_2021}. In the case of \gls{elliptical} galaxies, shapes are elongated along the external gravitational field~\citep{croft_weak-lensing_2000,catelan_intrinsic_2001}.
The shapes of \gls{spiral} galaxies are typically associated with their angular momentum, which arises from the \gls{torque} produced by the external gravitational field~\citep{heavens_intrinsic_2000, catelan_intrinsic_2001, codis_intrinsic_2015}. The alignments of spiral galaxy shapes are much weaker than for ellipticals and so far have not been directly observed \citep{zjupa_intrinsic_2020, johnston_kidsgama_2019, samuroff_dark_2022}. Some studies use galaxy spin instead of shapes to measure alignment \citep{lee_intrinsic_2011}, though here we focus on shapes as they can be more directly connected to \gls{cosmic shear} and are most commonly used in observational studies. \citet{lee_alignments_2007} provides a pedagogical overview of the physics and formalisms of spin alignments.

While IA can be used as a cosmological probe, historically it is most often studied as a contaminant of \gls{weak lensing}. 
As light travels to us from distant galaxies, it is bent by the gravitational field of the large-scale structure of the Universe and thus we observe distorted galaxy images.
When this effect is too small to be detected for individual galaxies, we say that we are in the weak lensing regime.
The resulting \gls{shear} in these galaxy shapes is known as \gls{cosmic shear}, and is a primary tool used to probe cosmological parameters \citep{Refregier_weak_2003, heymans_cfhtlens_2012, hildebrandt_kids-450_2017, hikage_cosmology_2019, abbott_dark_2023}.
The correlations between observed galaxy shapes used to measure weak lensing are difficult to separate from those that arise from IA.
Studies show that IA can account for a 30\% error on the matter power spectrum amplitude as measured by cosmic shear~\citep{hirata_intrinsic_2007}, making IA one of the most significant sources of \gls{systematic errors} in weak lensing measurements.

\headings{Additional note: galaxy types}\label{note:galaxy_types}

Throughout this guide, we refer to galaxies as ``early-type'' or ``late-type'', ``blue'' or ``red'', and ``elliptical'' or ``spiral'', depending on the reference we are following.
\Gls{early-type galaxies} are usually \gls{elliptical} or lenticular and tend to be redder in color. \Gls{late-type galaxies} are \gls{spiral} and typically blue.
The terminology ``early'' and ``late'' does not refer to the age of the galaxy, but to their ordering in the Hubble Sequence (also known as ``The Tuning Fork'') when Hubble initially thought that ellipticals evolve into spirals~\citep{hubble_extragalactic_1926}. Although sometimes used interchangeably, it is important to keep in mind that populations may be defined differently in different analyses.

\subsection{Reviews}
Here is a list of available reviews and primers on IA.
A Zotero group of key IA papers can be found below\footnote{\href{ https://www.zotero.org/groups/4989025/ia_key_papers}{zotero.org \slash groups\slash 4989025\slash ia\_key\_papers}}.

\begin{itemize}
    \item ``The intrinsic alignment of galaxies and its impact on weak gravitational lensing in an era of precision cosmology"~\cite{troxel_intrinsic_2015}\\
    \it{The first comprehensive review on intrinsic alignments, presenting extensive documentation of commonly used formalisms and the role of IA in precision cosmology.}

    \item ``Galaxy alignments: An overview"~\cite{joachimi_galaxy_2015}\\
    \it{Broad synopsis of IA, including physical motivations, a historical overview, and main trends.}

    \item ``Galaxy alignments: Observations and impact on cosmology"~\cite{kirk_galaxy_2015}\\
    \it{Descriptions of formalisms for measuring shapes and IA \glspl{tracer}, an overview of IA observations, and discussion of cosmological impacts and mitigation.}

    \item ``Galaxy alignments: Theory, modelling and simulations"~\cite{kiessling_galaxy_2015}\\
    \it{Detailed overview of common models and IA in $N$-body and \gls{hydrodynamic simulations}.}

\end{itemize}

\section{Ellipticity}\label{sec:Ellipticity}

\begin{figure*}[h]
\centering
\includegraphics[width=0.15\textwidth]{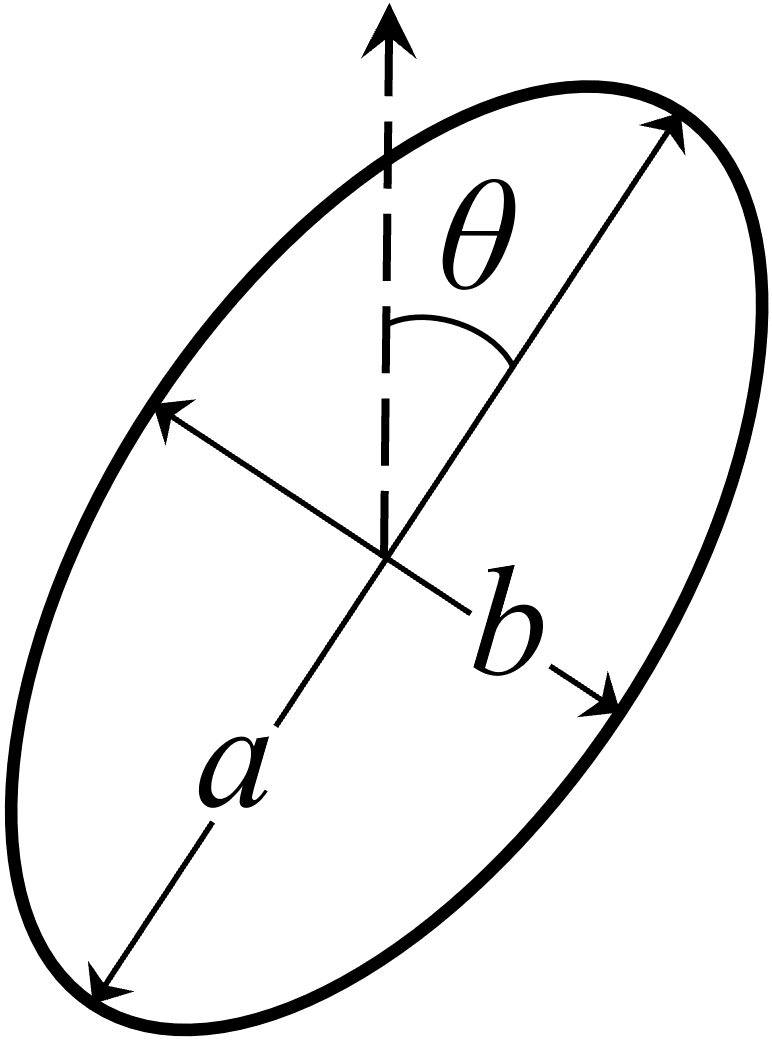}
\caption{The quantities $a$, $b$ and $\theta$ that define the shape and orientation of an ellipse. The dotted line indicates an arbitrary reference axis.}
\label{fig:ellipticity_definition}
\end{figure*}

IA studies model simulated, three-dimensional (3D) galaxy shapes as triaxial ellipsoids, and observed galaxies as their projected shape on the sky: two-dimensional (2D) ellipses.
This 2D \gls{ellipticity} is quantified in terms of the lengths of the major and minor axes of the ellipse ($a$ and $b$, respectively, with $b\le a$) and the orientation angle of the major axis, $\theta$, with respect to an arbitrary reference axis, as shown in \rfig{ellipticity_definition}.

For the detection of IA, ellipticity is typically measured relative to directions tracing the \gls{tidal field} (e.g., positions of galaxy overdensities in real data, or reconstructed tidal fields in simulations), often as a function of \gls{transverse} separation $r_p$.
By convention, the alignment signal is highest for very elongated shapes (larger axis ratio) that point along the direction of the tidal field.

While the formalisms below are standard, the methods of fitting shapes to observations vary across surveys and can impact the resulting IA signal.
The signal is also correlated with the clustering of the galaxy sample and depends on how far along the Line of Sight (\gls{LOS}) the measurement is averaged over.
Therefore, it is more common to use IA \glspl{correlation function} rather than a relative ellipticity, although ellipticity is a component of most \glspl{estimator}.

\subsection{Ellipticity: 2D Formalism}\label{sec:eformalism}

There are two different ways in which the \gls{ellipticity} of 2D shapes is commonly quantified. We will refer to these as $\varepsilon$ and $\chi$ and to ellipticity generically as $\epsilon$. In the rest of the document, $\epsilon$ can be taken to stand for either of the two ellipticity definitions.
These are defined as:
\begin{equation}
    \varepsilon = \frac{a-b}{a+b}\exp(2\mathrm{i}\theta)
\end{equation}
\begin{equation}
    \chi = \frac{a^2-b^2}{a^2+b^2}\exp(2\mathrm{i}\theta)\,.
\end{equation}
Both quantities are often referred to as the ellipticity, but $\chi$ is also known as the distortion \citep{mandelbaum_third_2014} or the normalized polarization \citep{viola_probability_2014}.
The definition $\varepsilon$ (but usually denoted $\epsilon$) is often used in \gls{weak lensing} studies because it is an unbiased \gls{estimator} of the \gls{cosmic shear}, $\gamma$. In contrast, $\chi$
must be adjusted by the responsivity $\mathcal{R}$ which quantifies the response of the ellipticity to an applied gravitational \gls{shear} ~\cite{bernstein_shapes_2002}:
\begin{equation}
\gamma = \frac{\langle\chi\rangle}{2\mathcal{R}}\,.\label{eq:responsivity}
\end{equation}
$\mathcal{R}= 1- \chi_\mathrm{rms}$ (rms is the root mean square) and is typically $\approx0.9$ depending on the galaxy sample~\citep{singh_intrinsic_2016}. In later parts of this guide, $\epsilon$ is used generally. It is assumed that where the $\chi$ definition is intended, it will have been adjusted for the responsivity so $\mathcal{R}$ is not explicitly shown.

\begin{figure*}
\centering
\includegraphics[width=\textwidth]{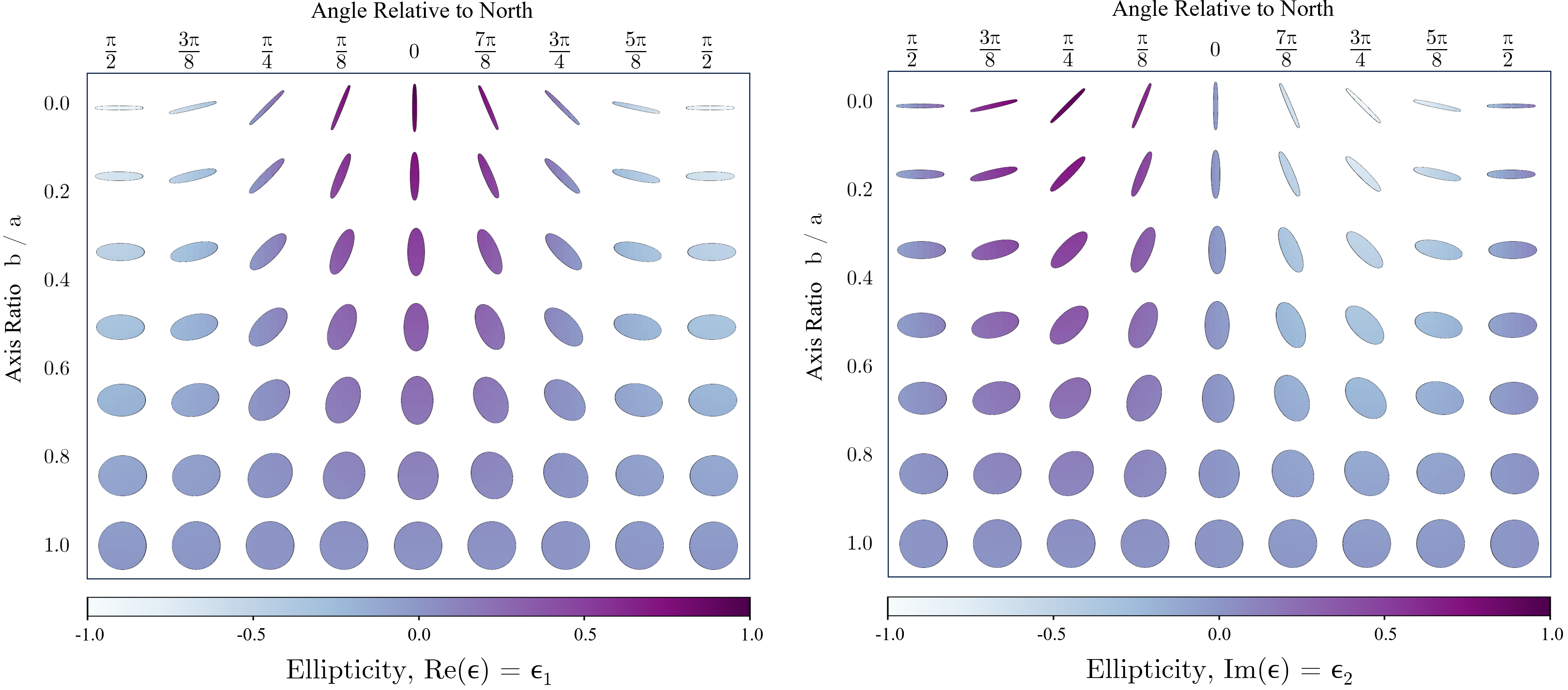}
\caption{Visualization of the real and imaginary components of \gls{ellipticity}, as described in Section~\ref{sec:Ellipticity}. These functionally contain the same information. 
$\epsilon_1$ is maximum when a shape is highly elongated and exactly aligned with the angle that the ellipticity is defined relative to, most commonly North. $\epsilon_2$ is maximum when the shape is aligned with $\pi/4$ away from the principal angle.}
\label{fig:ellipticity_visualizations}
\end{figure*}

Ellipticity is a complex quantity that can be broken up into its real and imaginary components, $\epsilon_1$ and $\epsilon_2$:
\begin{equation}\label{eq:ellipticity}
    \epsilon = \epsilon_1 + \mathrm{i}\epsilon_2\,,
\end{equation}
where
\begin{align}
   \epsilon_1 &= \lvert \epsilon\rvert\cos (2\theta)\\ 
 \epsilon_2 &= \lvert \epsilon\rvert\sin (2\theta)\,.
\end{align}
The factor of 2 arises because ellipticity is a spin-2 quantity, which means that it is invariant under rotations of integer multiples of $\pi$. See Figure~\ref{fig:ellipticity_visualizations}. 
The angle $\theta$ is usually defined as East of North. Its range can be $0-\pi$ or $\pm \frac{\pi}{2}$. 

$\epsilon_1$ represents the orientation of an ellipse relative to the direction where $\theta=0$ and $\epsilon_2$ represents the orientation relative to the direction where $\theta=\frac{\pi}{4}$.
Note that $\epsilon_1$ and $\epsilon_2$ contain the same information (Figure~\ref{fig:ellipticity_visualizations}).
When measured relative to another galaxy or the direction of the \gls{tidal field}, they are usually denoted as $\epsilon_+$ and $\epsilon_\times$, with the subscripts respectfully read as ``plus" and ``cross".
$\epsilon_+ > 0$ indicates an alignment along the tidal field direction, and $\epsilon_+<0$ indicates a tangential orientation as seen in gravitational \gls{shear} (\rsec{shear}).
$\epsilon_\times$ is equivalent to $\epsilon_2$.
On average, $\epsilon_\times$ is 0 for galaxies in an isotropic Universe. 

\subsection{Ellipticity: Modeling and 3D Formalism}

\headings{Ellipticity from particles}

Simulated galaxies and \glspl{halo} are usually modeled as triaxial ellipsoids composed of ``particles''.
To summarize the shape of these objects, it is common to use the inertia tensor, I, which is computed by summing over the positions $i,j \in (1,2,3)$ of $N$ particles.
To better approximate the object's position and shape, this sum is usually weighted by particle mass or luminosity (when applicable).
For particles $k$ with weights $w_k$ that sum to $W$, this form of the moment of inertia tensor is~\citep{samuroff_advances_2021}
\begin{equation}\label{eq:inertia_tensor}
    I_{ij} = \frac{1}{W}\sum^N_{k=1} w^kx_i^kx_j^k\,.
\end{equation}
It is also common to weight $I_{ij}$ by distance from the center of the object.
This center-weighting produces the reduced inertia tensor \citep{chisari_intrinsic_2015},
\begin{equation}\label{eq:reduced:inertia_tensor}
\tilde{I}_{ij} = \frac{1}{W}\sum^N_{k=1} w^k\frac{x_i^kx_j^k}{r_k^2}\,.
\end{equation}
$r_k$ is the distance between the particle $k$ to the object's center of mass. The reduced inertia tensor can provide a better approximation of the shape of the object at its center~\citep{joachimi_intrinsic_2013}. Equation \ref{eq:reduced:inertia_tensor} is known to create a coherent bias in shapes, since the $r$ weighting kernel is inherently round. To avoid this limitation, studies often calculate the inertia tensor on particles within a fixed volume, while iteratively rescaling the axis lengths until they converge~\citep{schneider_shapes_2012, mandelbaum_great3_2015}.

An ellipsoid can be constructed using the eigenvectors and eigenvalues of $I_{ij}$, which can then be projected along the $z$-axis into its 2D second moments $Q_{11}$, $Q_{22}$, $Q_{12}+Q_{21}$ (Eq. 5.37 of~\cite{bartelmann_weak_2001}). The projected \gls{ellipticity} $\epsilon$ is obtained via
\begin{equation}
    \epsilon = \frac{(Q_{11} - Q_{22}, 2Q_{12})}{Q_{11}+Q_{22}+2\sqrt{\det\mathbf{Q}}}\,
\end{equation}
where $\mathbf{Q}\equiv Q_{ij}$ with $i,j\in(1,2)$. Note that this assumes that ellipticity is defined by $\varepsilon$ from \rsec{eformalism}. If the ellipticity is instead defined as $\chi$ (distortion or normalized polarization), the denominator is simply $Q_{11}+Q_{22} $ - see \rsec{eformalism} and~\cite{mandelbaum_third_2014}.

\begin{figure*}
\centering
\includegraphics[scale=0.4]{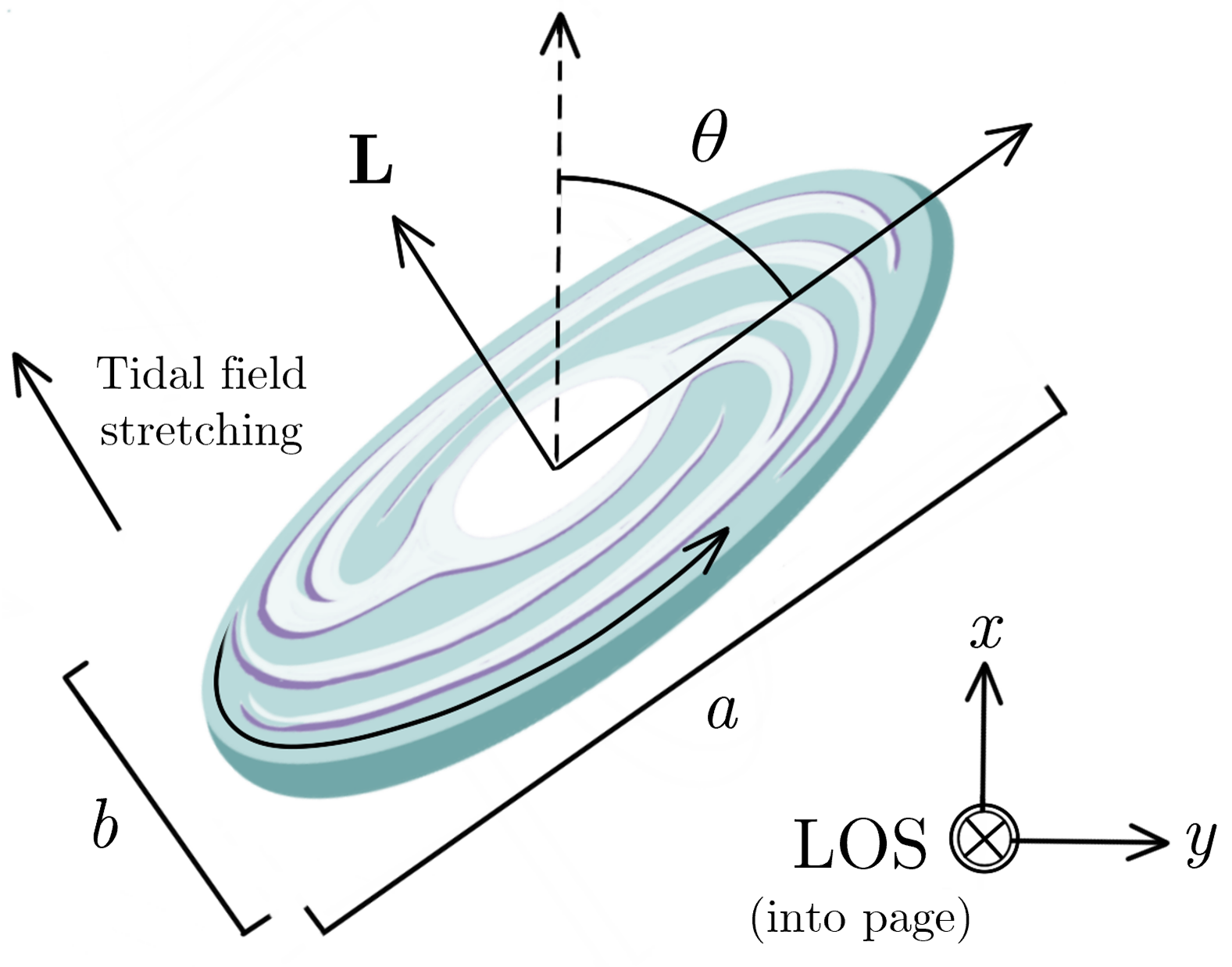}
\caption{The setup for obtaining the projected shape of a \gls{spiral} galaxy. In linear theory, the angular momentum vector $\mathbf{L}$ of the galaxy is aligned along the direction of tidal stretching. The projected axis ratio, $b/a$, is a function of $\mathbf{L}$ and the ratio of the disk's intrinsic thickness to its diameter (not shown here).\vspace{.2in}}
\label{fig:projecting_spiral}
\end{figure*}

\headings{Projected Ellipticity from angular momentum}

\Gls{late-type galaxies} are typically modeled as circular discs and their \gls{ellipticity} is often assumed to be aligned with their angular momentum (tidal torquing or spin alignment), $\mathbf{L} = \left\{L_x, L_y, L_{\parallel}\right\}^{\tau}$. $\tau$ denotes the transpose vector(Figure~\ref{fig:projecting_spiral}).
To obtain the projected shape of a \gls{spiral} galaxy along the \gls{LOS}, or $ L_{\parallel}$, the orientation angle $\theta$ is given by:
\begin{equation}
    \theta = \frac{\pi}{2} + \text{arctan}\left(\frac{L_y}{L_x}\right)\,,
\end{equation}
and the axis ratio:
\begin{equation}
    \frac{b}{a} = \frac{\lvert L_\parallel \lvert}{\lvert\mathbf{L}\lvert} + r_{\text{edge-on}} \sqrt{1 - \frac{L_\parallel^2}{\lvert \mathbf{L}\rvert^2}}\,.
\end{equation}
$r_{\text{edge-on}}$ describes the ratio of the disc thickness to disc diameter, which is approximately equivalent to the axis ratio for a galaxy viewed edge-on; this contribution is expected to be significant for galaxies with bulges~\citep{joachimi_intrinsic_2013}.
Assuming linear tidal torquing, a \gls{halo}'s spin is written as~\citep{lee_nonlinear_2008}
\begin{equation}
    L_i \propto \epsilon_{ijk} I^k_l T^{jl}\,,
    \label{eq:tidal_torquing}
\end{equation}
where ${i,j,k}={1,2,3}$ in the three spatial directions, $\epsilon_{ijk}$ the \gls{Levi-Civita symbol} and $T^{jl}$ the gravitational tidal \gls{shear}.
The latter, which is a symmetric tensor, is defined as~\cite[e.g.][]{blazek_testing_2011}
\begin{equation}\label{eq:tidal_shear}
    T_{ij} = \frac{\partial^2 \Phi}{\partial x_i \partial x_j}\,,
\end{equation}
where $\Phi$ is the gravitational potential, $x_{i,j}$ represents comoving Cartesian coordinates, and the indices $\{i,j\}=\{1,2,3\}$ indicate the three spatial directions.
Following \req{tidal_torquing}, tidal torquing leads to quadratic alignments of galaxy shapes with the tidal shear.

\newpage
\headings{Ellipticity from tidal field}

\begin{figure}
\centering
\includegraphics[width=8cm]{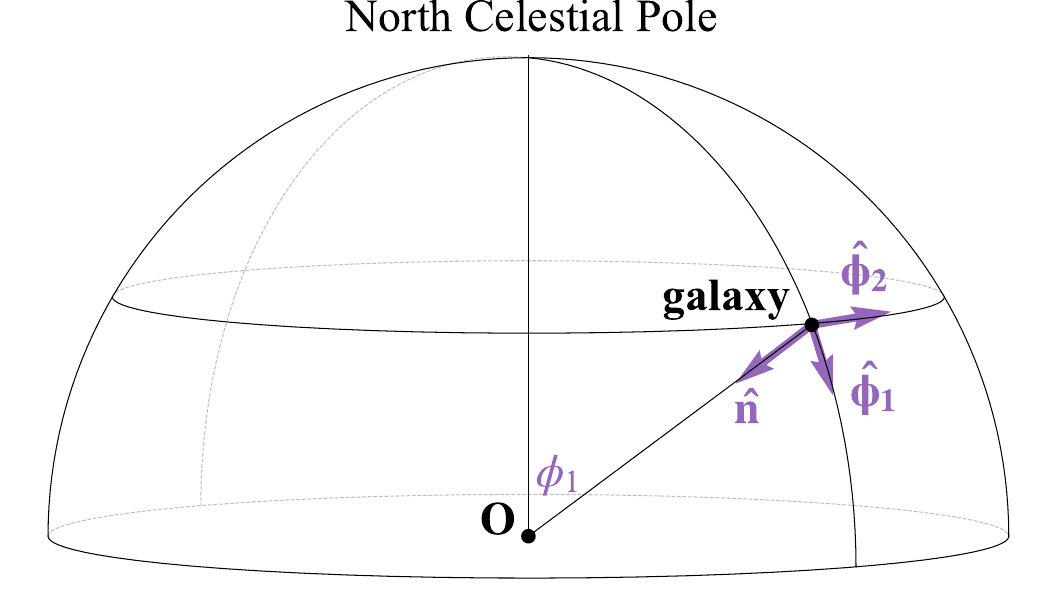}
\caption{Representation of the Northern Celestial Hemisphere.
The observer is indicated by {\bf O} and a given observed galaxy by {\bf galaxy}.
The purple basis indicates the rotated basis defined in \req{rotated_basis_1} and \req{rotated_basis_2}, whereas $\phi_1$ is the angle from the North Pole. \vspace{.2in}}
\label{fig:3D_basis}
\end{figure}

\Gls{early-type galaxies} are considered to be triaxial ellipsoids whose axes align with the underlying gravitational tidal \gls{shear}, $T_{ij}$~\citep{catelan_intrinsic_2001}.
In order to derive the predicted galaxy ellipticities given $T_{ij}$, we can project the 3D \gls{tidal field} along two axes at the location of each galaxy.
The convention is to project along the galaxy's North Pole distance, $\phi_1$, and right ascension, $\phi_2$ (see~\rfig{3D_basis}).
The latter is the angle complementary to declination. In this setting, $\epsilon_1>0$ corresponds to east-west elongation and $\epsilon_2>0$ corresponds to northeast-southwest elongation.
We first consider a Cartesian orthonormal basis, $(\hat{\mathbf{x}},\hat{\mathbf{y}},\hat{\mathbf{z}})$, at the location of each galaxy. We then rotate this basis into $(\mathbf{\hat{n}},\boldsymbol{\hat{\phi}_1},\boldsymbol{\hat{\phi}_2})$, such that $\mathbf{\hat{n}}$ is parallel to the \gls{LOS} to the galaxy.
These two bases are related by
\begin{eqnarray}
\label{eq:rotated_basis_1}
\boldsymbol{\hat{\phi}_1} &=& \cos{\phi_1}\cos{\phi_2}\,\hat{\mathbf{x}}+ \cos{\phi_1}\sin{\phi_2}\,\hat{\mathbf{y}} - \sin{\phi_1}\,\hat{\mathbf{z}}\,,\\
\label{eq:rotated_basis_2}
\boldsymbol{\hat{\phi_2}} &=& -\sin{\phi_2}\,\hat{\mathbf{x}}+\cos{\phi_2}\,\hat{\mathbf{y}}\,.
\end{eqnarray}
The next step is to intermediately define the linear combinations,
\begin{eqnarray}
    m^i_{+} &=& \frac{1}{\sqrt{2}}\left(\boldsymbol{\hat{\phi}_2^i} - i \boldsymbol{\hat{\phi}_1^i}\right)\\
    m^i_{-} &=& \frac{1}{\sqrt{2}}\left(\boldsymbol{\hat{\phi}_2^i} + i \boldsymbol{\hat{\phi}_1^i}\right)\,.
\end{eqnarray}
Having defined the above fields, we can decompose the 3D tidal \gls{shear} into the rotated basis, as~\citep{schmidt_cosmic_2012}
\begin{equation}
    T_\pm = \sum_{i=1}^3\sum_{j=1}^3 {m^i}_{\mp} {m^j}_{\mp} T_{ij}\,,
\end{equation}
where $T_\pm$ are the 2D ellipticities, such that~\citep{tsaprazi_field-level_2022}
\begin{equation}
    \epsilon_1 \pm i\epsilon_2 = - \frac{C_1}{4\pi G} T_\pm\,,
\end{equation}
in the nonlinear alignment model, for example, described in \rssec{NLA}.
Here $C_1$ is the IA amplitude (introduced in \ref{subsec:lin_alignment_model}) and $(\epsilon_1,\epsilon_2)$ the observed galaxy ellipticities.

\subsection{Ellipticity: Additional Notations}\label{sec:ellipticity_addnotation}

\begin{itemize}

    \item $\epsilon$: frequently used rather than $\varepsilon$ and sometimes instead of $\chi$.
    
    \item $e$: sometimes used equivalently to $\epsilon$ or $\chi$.
    
    \item $\epsilon_\text{T}$ or $\epsilon_\text{t}$: the tangential component of ellipticity, equivalent to $\epsilon_+$.
    \item $\eta$: sometimes used instead of $\chi$.
    
    
    \item $\phi$: sometimes used for the orientation angle.

    \item $a,b$: sometimes denote the length of the semi-axes of the ellipse, rather than the full axes lengths.

    \item $q$: the ratio of the ellipse axes, $b/a$, with $b\le a$.
\end{itemize}

\newpage
\subsection{Ellipticity: References}

\begin{itemize}
    \item ``On the intrinsic shape of elliptical galaxies"~\cite{binggeli_intrinsic_1980}\\
    \it{Historical discussion of the true and projected shapes of \gls{elliptical} galaxies}.

    \item ``Weak gravitational lensing"~\cite{bartelmann_weak_2001}\\
    \it{Review paper, Section 4 contains additional ellipticity details}.

    \item ``Shapes and shears, stars and smears: Optimal measurements for Weak Lensing"~\cite{bernstein_shapes_2002}\\
    \it{Defines the \gls{shear} responsivity factor, $\mathcal{R}$.}
    
    \item ``Means of confusion: How pixel noise affects shear estimates for weak gravitational lensing"~\cite{melchior_means_2012}\\
    \it{Appendix A discusses the pros and cons of different ellipticity definitions}.
    
    \item ``Intrinsic galaxy shapes and alignments I: Measuring and modelling COSMOS intrinsic galaxy ellipticities"~\cite{joachimi_intrinsic_2013}\\
    \it{Contains condensed formulae for projecting 3D shapes to 2D}.
    
    \item ``The third gravitational lensing accuracy testing (GREAT3) challenge handbook"~\cite{mandelbaum_third_2014}\\
    \it{Section 2.1 includes different ellipticity definitions}.

    \item ``Intrinsic alignments of galaxies in the Illustris simulation"~\cite{hilbert_intrinsic_2017}\\
    \it{Additional notation and definitions for 3D shapes}.

    \item ``The mass dependence of dark matter halo alignments with large-scale structure"~\cite{piras_mass_2018}\\
    \it{Derivations for the 3D shapes of galaxies from the moment of inertia tensor}.
    
    \item ``Galaxy shape statistics in the effective field theory"~\cite{vlah_galaxy_2021}\\
    \it{Formalism for projecting 3D shapes in forms that are convenient for numerical implementation}.
    \item ``Intrinsic alignment as an RSD contaminant",~\cite{ lamman_intrinsic_2023}\\
    \it{Appendix A contains condensed relations for flat projection of a triaxial shape}.

\end{itemize}

\begin{figure}[ht] 
\begin{center}
\includegraphics[scale=0.45]{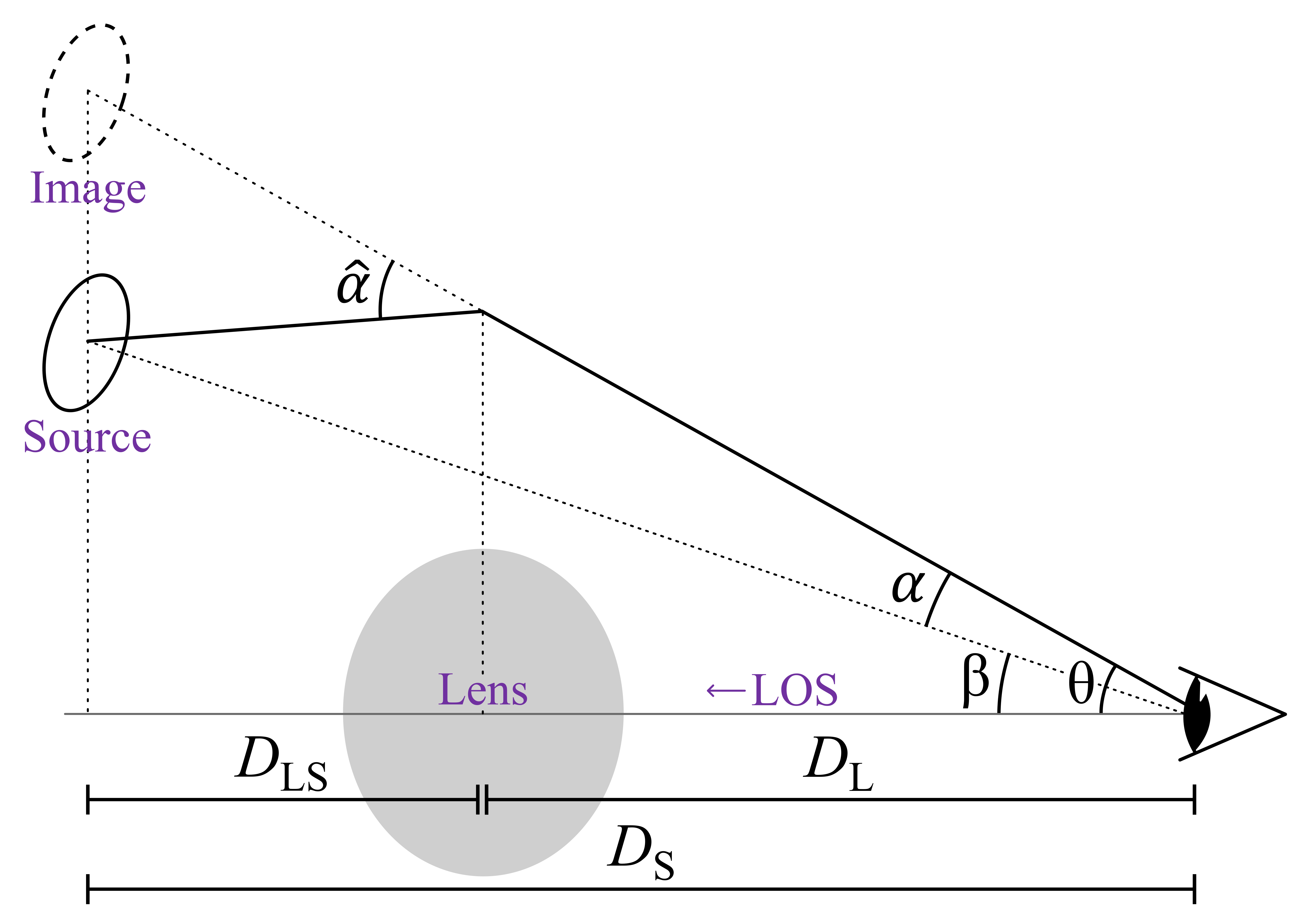}
\end{center}
\caption{A diagram of a gravitational lensing system.
A mass concentrated at an angular diameter distance $D_{\text{L}}$ from the observer lenses the light from a source located at distance $D_{\text{S}}$.
In the \gls{thin-lens approximation}, the true angular position of the light source $\boldsymbol{\beta}$ with respect to the \gls{LOS} is related to the observed angular position $\boldsymbol{\theta}$ through the lens or ray-tracing equation, $\boldsymbol{\beta} = \boldsymbol{\theta} - \boldsymbol{\alpha}(\boldsymbol{\theta})$, where $\boldsymbol{\alpha}$ is the reduced deflection angle measured by the observer. The angle $\hat{\boldsymbol{\alpha}}$ is the deflection angle measured at the lens.
Note that angles are exaggerated to aid visualization.\vspace{.2in}}
\label{fig:lensing_diagram}
\end{figure}

\section{Shear}\label{sec:shear}

IA is often measured as a contaminant of \gls{cosmic shear}~\citep{bernstein_shapes_2002, hirata_intrinsic_2004} and is sometimes referred to as \textit{intrinsic shear}, $\gamma_{\text{I}}$.
The coherent distortion of galaxy light by foreground mass creates a tangential \gls{shear} on the sky, $\gamma_\text{lensing}$, often acting in opposition to the signal from \gls{tidal alignment} at large scales.
As these two phenomena are difficult to distinguish observationally and are necessarily measured together, much of the IA formalism described in this and the following sections is from \gls{weak lensing}.

\subsection{Shear: Formalism}

The observed \gls{shear} signal is the combination of the intrinsic component of shapes and the component gravitationally lensed by foreground mass:
\begin{equation}
    \gamma_{\text{observed}} = 
    \gamma_{\text{I}} +
     \gamma_{\text{lensing}}\,.
\end{equation}
This mathematical sum is an approximation for when the effects are small, as in \gls{weak lensing}; see the note about \hyperref[note:addition]{shear addition} below. 
Shear quantifies the shape of galaxies and thus is related to the \gls{ellipticity} described in \rsec{eformalism}. 
Lensing is often defined via the critical mean density, $\Sigma_\text{crit}$, which is the maximum surface density before the light of a source is split into multiple images by a foreground mass.
 $\Sigma_\text{crit}$ is a function of fundamental constants and the \gls{radial} separations involved in the lensing system, as illustrated in \rfig{lensing_diagram}.
\begin{equation}
    \Sigma_\text{crit} = \frac{c^2 D_\mathrm{S}}{4\pi G D_\mathrm{L} D_\mathrm{LS}}.
\end{equation}
For instance, while \gls{cosmic shear} is a direct measurement of the large-scale structure between the source and observer, \gls{galaxy-galaxy lensing} is the cross-correlation between source galaxies and biased \glspl{tracer} of the underlying matter~\citep{sheldon_galaxy-mass_2004, heymans_kids-1000_2021, prat_dark_2022}. 
It is the difference between the average surface density of galaxies within some projected separation $\bar{\Sigma}(<r_p)$, and the surface density at separation $\Sigma(r_p)$, as a fraction of $\Sigma_{\rm crit}$:
\begin{equation}
    \gamma_{\text{overdensity}} = \frac{\bar{\Sigma}(<r_p) - \Sigma(r_p)}{\Sigma_{\rm crit}}\,.
\end{equation}

\headings{Variable definitions}
\begin{itemize}
    \item $\gamma$: total shear, as described in \rsec{correlations}.
    
    \item $\gamma_\mathrm{I}$: intrinsic shear due to \gls{tidal alignment}.
    
    \item $\langle\epsilon\rangle$: expectation value of the ellipticity.
    
    \item $\Sigma$: surface overdensity, projected to the plane of the sky, i.e., the mass density integrated along the \gls{LOS}.
    
    \item $D_\mathrm{S}$: \gls{radial} distance between observer and light source.
    
    \item $D_\mathrm{L}$: radial distance between observer and gravitational lens.
    
    \item $D_\mathrm{LS}$: radial distance between light source and lens.
    
    \item $c$: speed of light.
    
    \item G: gravitational constant.
\end{itemize}

\subsection{Shear: Additional Notation}

\begin{itemize}
    \item $\gamma_t$ or $\gamma_T$ or $\gamma_+$: tangential \gls{shear}.
    $\langle\gamma_\times\rangle =0$ in the absence of systematics, like $\epsilon_+$, so it is often assumed that $\gamma_t = \gamma$.
    
    \item $\gamma_\mathrm{IA}$: alternative notation for $\gamma_\mathrm{I}$, intrinsic shear due to \gls{tidal alignment}.
\end{itemize}

\headings{Additional note: shear addition}\label{note:addition}

Since \gls{shear} matrices can be asymmetrical, they do not form a group under matrix multiplication, and therefore shear terms are not commutative under matrix addition.
Instead, we can define a group with an addition operation,
\begin{equation}
    \boldsymbol{\gamma}_3 = \boldsymbol{\gamma}_2 \oplus \boldsymbol{\gamma}_1 \quad \Longleftrightarrow \quad \mathbf{S}\boldsymbol{\gamma}_3\mathbf{R} = \mathbf{S}\boldsymbol{\gamma}_2\mathbf{S}\boldsymbol{\gamma}_1\,,
\end{equation}
where $\mathbf{S}_\gamma$ is the shear matrix, and $\mathbf{R}$ is the unique rotation matrix that allows $\mathbf{S}\boldsymbol{\gamma}_3$ to be symmetric.
For \gls{weak lensing} in general, the shears are assumed to be small, such that using the mathematical addition is a valid approximation. 
However, this is not always true for IA. 
We refer the reader to~\cite{miralda-escude_correlation_1991} and~\cite{bernstein_shapes_2002}
for the derivation of the appropriate addition formalism.

\headings{Additional note: convergence}\label{note:convergence}

\begin{figure}[h]  
\begin{center}
\includegraphics[scale=0.45]{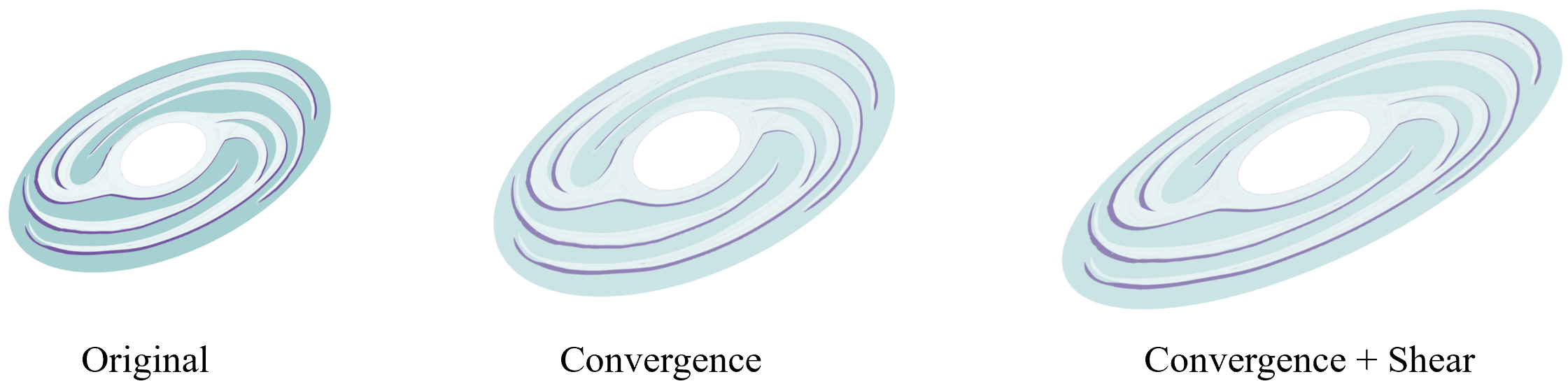}
\end{center}
\caption{A cartoon demonstrating the difference between \gls{convergence} and \gls{shear}.
Convergence isotropically changes the apparent size of a galaxy while conserving surface brightness, demonstrated here by a fainter color.
Shear elongates the galaxy.}
\label{fig:convergence_cartoon}
\end{figure}

\Gls{weak lensing} has two effects on observed galaxies: \gls{shear},  $\gamma$, and \gls{convergence}, $\kappa$, shown in Figure~\ref{fig:convergence_cartoon}.
The shear is \gls{trace-free} and characterizes the anisotropic stretching of the galaxy's source image by quantifying the projection of the gravitational \gls{tidal field}. Convergence measures the surface mass density and is an isotropic distortion, describing the change in the size of the lensed galaxy while maintaining a constant surface brightness.

When galaxy shapes are measured, we actually measure the reduced shear,
    \begin{equation}
        g = \frac{\gamma}{1-\kappa}\,,
    \end{equation}
since we measure shapes, not sizes.
This is an invariant quantity that introduces a \gls{mass-sheet degeneracy}~\citep{schneider_weak_2008}.
Since $\kappa\ll 1$ in the \gls{weak lensing} regime, the \gls{shear} is a good approximation of the reduced shear.
Note that $g$ in the above equation is not to be confused with the variable g that is used to describe galaxy positions, nor with the overdensity.
Both shear and \gls{convergence} contribute to the magnification, $\mu$, which is defined as the ratio of lensed to unlensed flux:
\begin{equation}
    \mu = \frac{1}{(1-\kappa)^2-|\gamma|^2}\,.
\end{equation}
In the weak lensing regime, $\mu\approx 1+2\kappa$. Magnification impacts both the apparent position of galaxies and the distribution of light received from a single galaxy: in areas with positive (negative) convergence, the apparent distance between any two objects on a source plane is increased (decreased), and the captured fraction of the solid angle of light emitted from a source is amplified (reduced). Magnification thus affects the observed area number density and the selection probability of individual galaxies, thereby impacting the observed number density of objects in large-scale structure surveys.

\subsection{Shear: References}

\begin{itemize}
    \item ``A method for weak lensing observations"~\cite{kaiser_method_1995}\\
    \it{Describes the motivation for measuring \gls{shear} and how to model the shear response}.
    
    \item ``Weak gravitational lensing"~\cite{bartelmann_weak_2001}\\
    \it{Comprehensive review of all aspects of \gls{weak lensing}.}
    
    \item ``Cosmology with cosmic shear observations: a review"~\cite{kilbinger_cosmology_2015}\\
    \it{More recent review of all aspects of cosmic shear.}
    \item ``Weak gravitational lensing"~\cite{bartelmann_weak_2017}\\
    \it{Basic introduction to concepts, assuming little prior knowledge.}

    \item ``Weak Lensing for Precision Cosmology"~\cite{mandelbaum_weak_2018}\\
    \it{Review of weak lensing in modern cosmology.}
\end{itemize}

\section{IA Correlation Function Notation}\label{sec:correlations}

The observed shape-density and shape-shape correlations are the result of several combinations of effects, including physical galaxy correlations and lensing. This section describes the notation most commonly used to denote these effects, while \rsec{IACF} describes methods to quantify their correlations.

\begin{figure}[h]  
\begin{center}
\includegraphics[scale=0.5]{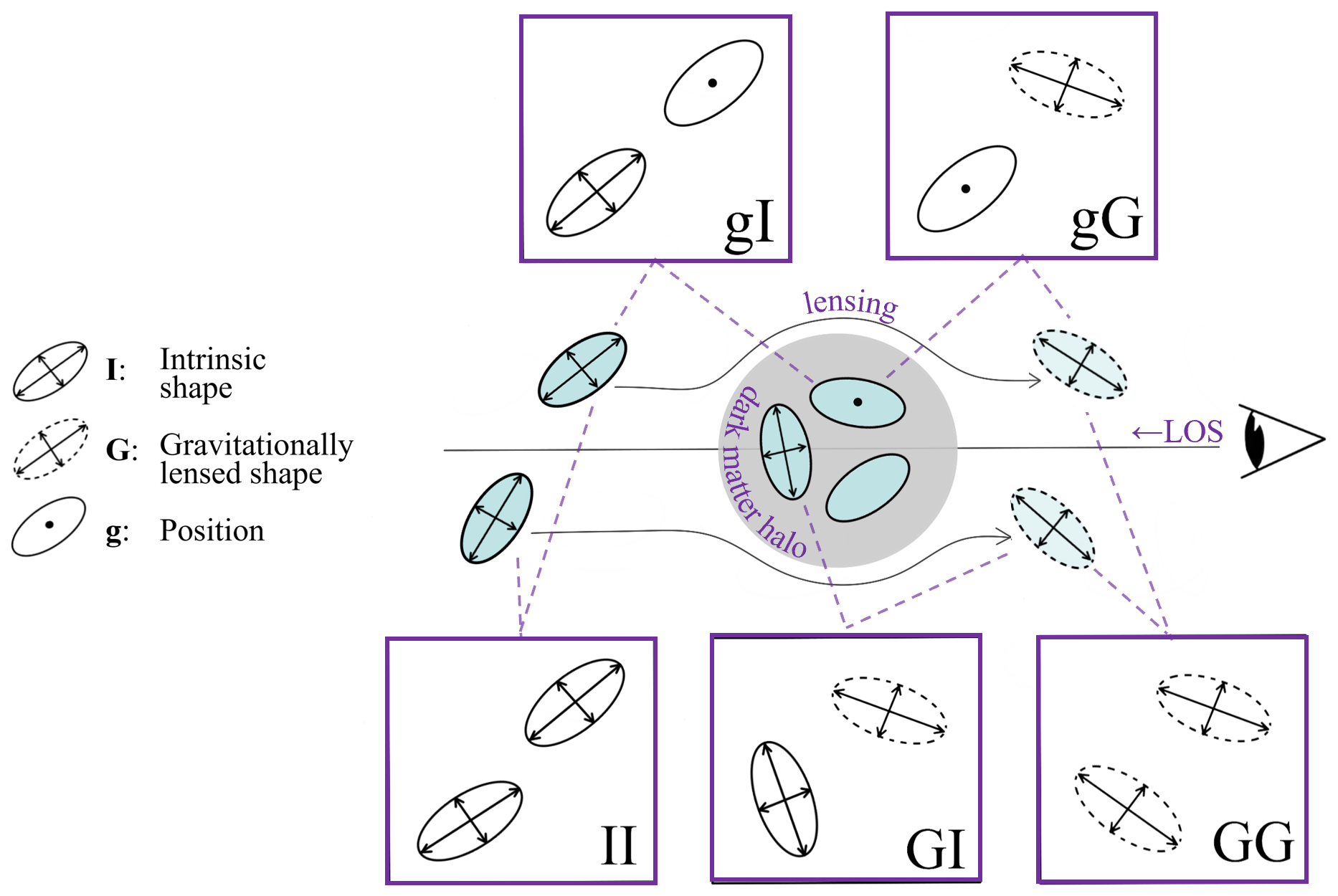}
\end{center}
\caption{A schematic of correlations relevant to IA and \gls{weak lensing} studies, shown along the \gls{LOS}. 
Here, two background galaxies are lensed by a foreground dark matter \gls{halo}.
The observed shape $\epsilon$ of a galaxy is a combination of its intrinsic (I) and lensed (G) components.
These are correlated with the underlying dark matter density, which can be traced by galaxy shapes and positions (g).
Correlations are represented by combining these notations.
For example, gI is the correlation between galaxy positions and intrinsic galaxy shapes, and GI is the correlation between the lensed component of galaxy shapes and intrinsic galaxy shapes. GI correlations are important because overdensities cause IA between galaxies at the same \gls{redshift} while also lensing more distant galaxies. Separating galaxies into widely-spaced tomographic bins can distinguish between the GI and II terms (Section~\ref{note:tomography},  Figure~\ref{fig:tomography_diagram}). For other helpful diagrams of IA correlations, see Figure 1.6 of~\cite{Fortuna_galaxy_2021} and Figure 6 of~\cite{troxel_intrinsic_2015}. \\
\textit{Note that this is a cartoon; galaxy shapes, orientations, and positions are only symbolic.
These correlations are never measured for individual galaxies and are typically only statistically significant when measured for at least $10^{4}$ objects.}}
\label{fig:correlations_diagram}
\end{figure}

\subsection{Correlations: Formalism}

The main correlated quantities relevant to IA are the intrinsic shape of galaxies, the component of the shape that is gravitationally lensed, and the position of galaxies (\rfig{correlations_diagram}).
For observed correlations that are measured as a function of sky separation, these are most commonly notated as
\begin{itemize}
    \item I: intrinsic galaxy shape.

    \item G: lensed component of shape, also referred to as ``extrinsic" shape.

    \item g: galaxy position (used as a \gls{tracer}).
\end{itemize}
The observed shape-shape correlation $\langle \epsilon_i\epsilon_j \rangle$ between galaxies in two \gls{radial} bins, $i$ and $j$, is the sum of every G and I combination: 
\begin{eqnarray}
    \langle \epsilon_i \epsilon_j \rangle = \langle \text{G}_i\text{G}_j \rangle + \langle \text{G}_i\text{I}_j \rangle + \langle \text{I}_i \text{G}_j \rangle + \langle \text{I}_i\text{I}_j \rangle\,. \label{eq:correlation}
\end{eqnarray}
These effects do not necessarily sum mathematically (see the note above on \hyperref[note:addition]{shear addition}). The notation $\langle X_iY_j \rangle$ indicates the correlation of a quantity $X$ of a sample $i$ relative to quantity $Y$ of sample $j$ (\rssec{corrfunc_form}).
For example, $\langle \text{G}_i \text{I}_j \rangle$ is the correlation between gravitational \gls{shear} of one sample relative to the intrinsic shapes of another sample.
For binning along the \gls{LOS} (see Section~\ref{note:tomography} on tomography), if $i$ is in a closer \gls{radial} bin than $j$ then $\langle \text{G}_i \text{I}_j \rangle$ will be 0 for non-overlapping \gls{redshift} bins since lensed shapes are created by a foreground mass.
Therefore, either the second or the third term of \req{correlation} will be equal to zero.
The first term of \req{correlation} is the \gls{shear} correlation. This contains the majority of cosmological information, but cannot be directly measured because the observed signal includes the IA terms.
The last term is the correlation between intrinsic ellipticities.
Similarly, the observed galaxy shape-density correlation $\langle \epsilon_i n_j \rangle$ contains contributions from cross terms between lensed shape and density, and intrinsic shape and density,
\begin{eqnarray}\langle \epsilon_i n_j \rangle = \langle \text{G}_i n_j \rangle + \langle \text{I}_i n_j \rangle\,.
\end{eqnarray}
The observed galaxy number density is the sum of the intrinsic number density g and a lensing magnification component $m$ due to foreground overdensities (see the note above on \hyperref[note:convergence]{convergence}). This expression can then be written as
\begin{eqnarray}\langle \epsilon_i n_j \rangle = \langle \text{G}_i \text{g}_j \rangle + \langle \text{I}_i \text{g}_j \rangle\ + \langle \text{G}_i m_j \rangle\ + \langle \text{I}_i m_j \rangle\,,
\end{eqnarray}
where $\langle \text{G}_i \text{g}_j \rangle$ is usually called the \gls{galaxy-galaxy lensing} signal. See~\cite{joachimi_simultaneous_2010} for a cosmological analysis using a joint treatment of galaxy \gls{ellipticity}, galaxy number density, and their cross-correlations.

\subsection{Correlations: Additional Notations}

\begin{itemize}
    \item $+$: As defined in \rsec{eformalism}, the component of shape relative to the direction of the \gls{tidal field}.

    \item $\times$: As defined in \rsec{eformalism}, the component of shape relative to $\frac{\pi}{4}$-off the direction of the tidal field.

    \item $T$ or $t$: the ``tangential" component of shape, equivalent to $+$~\citep{hilbert_intrinsic_2017}.

    \item $\gamma_\text{obs}$: total observed shape, similar to $\epsilon$.

    \item $\gamma$: gravitational \gls{shear}, or the correlation between lensed shapes. Also sometimes used as the total observed shape.
    $\langle\gamma\gamma\rangle$ is equivalent to GG.

    \item $\delta_g = \frac{g-\bar{g}}{\bar{g}}$: fractional density of galaxies, often used interchangeably with g.

    \item $\delta_{\text{m}}$: fractional density of matter, as opposed to overdensity traced by galaxies.
    Sometimes notated as just $\delta$.


    \item $m$: sometimes used to refer to magnification from lensing, see Section~\ref{note:convergence}.

    \item $\eta_e$: correlation between the 3D shape of a galaxy and the position of another galaxy~\citep{tenneti_intrinsic_2015, chisari_intrinsic_2015}.
    This is different from the lensing parameter in \rsec{shear}.

    \item $\eta_s$: correlation between the spin direction of a galaxy and position of another galaxy~\citep{chisari_intrinsic_2015}.
\end{itemize}

\begin{figure}[h]  
\begin{center}
\includegraphics[scale=0.34]{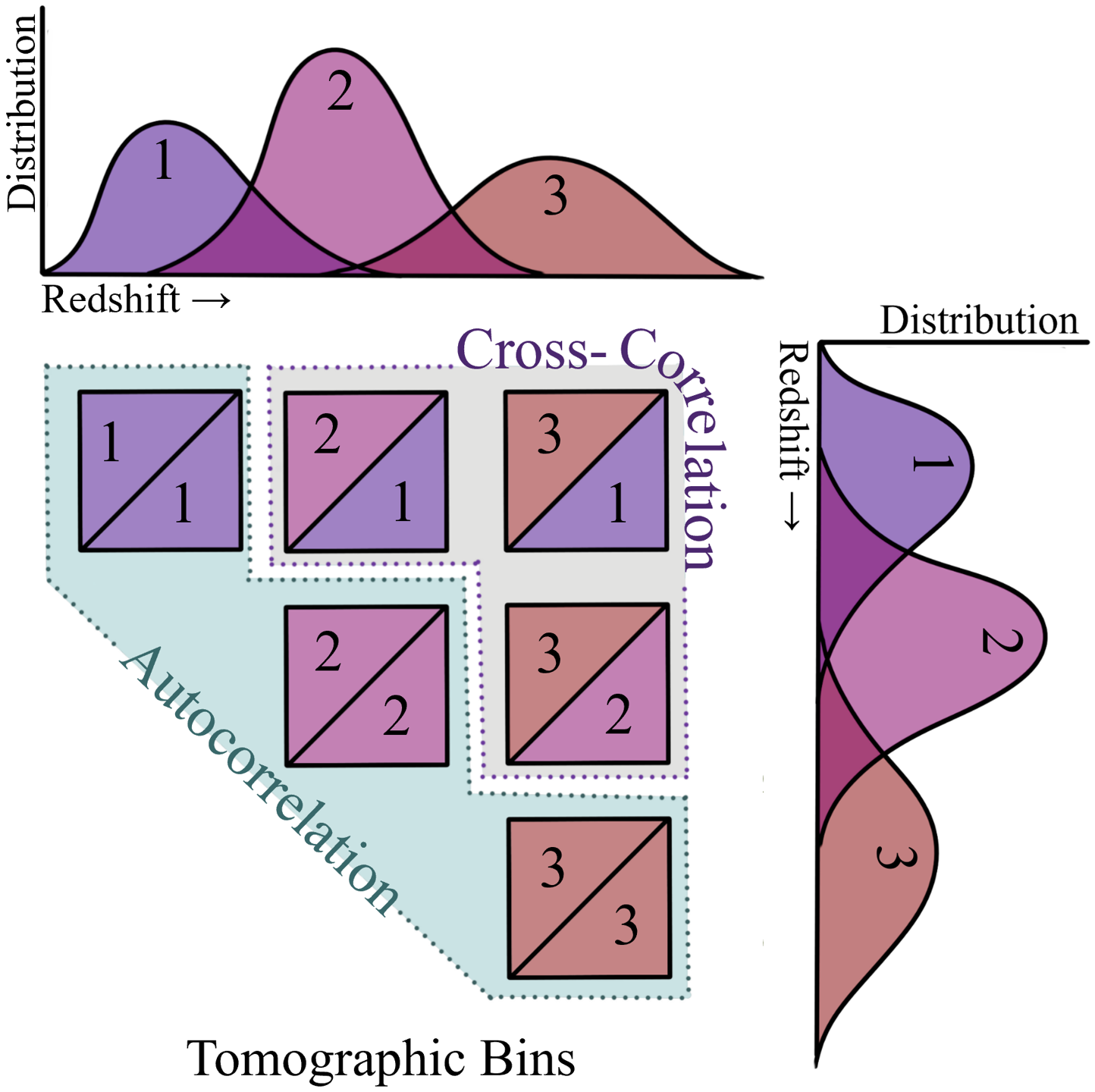}
\end{center}
\caption{An illustrative example of binning used to measure \gls{cosmic shear}.
The initial samples are sliced into three tomographic bins by \gls{redshift} (real surveys typically use more).
The redshift distributions shown on the vertical and horizontal axes can be from the same or different samples.
The boxes are number- and color-coordinated to show the six different combinations for correlations.
If the correlations are done within the same bin, they are referred to as auto-correlations.
Correlations between different bins are called cross-correlations.}
\label{fig:tomography_diagram}
\end{figure}

\headings{Additional note: tomography}\label{note:tomography}

Tomography in the context of cosmological correlations refers to the technique where the \gls{redshift} distribution of galaxies, $n(z)$, is sliced into redshift bins, also referred to as tomographic bins (Figure \ref{fig:tomography_diagram}).
This method allows the extraction of additional information from correlations within the galaxy sample that would otherwise be projected out, for example the growth of structure as a function of time.
Binning can be done so that bins are equally separated in redshift, so that there is an equal number of galaxies in each bin or some combination of the two.
If correlations are measured inside one bin, they are called ``auto-correlations".
Correlations between galaxies in different tomographic bins are referred to as ``cross-correlations".
Both \gls{spectroscopic} and \gls{photometric} surveys utilize the tomographic technique.
In spectroscopic surveys, it is possible to avoid the overlapping of redshift distributions between bins.
This is not possible in photometric surveys because the redshifts are not known precisely or accurately enough. For more information on the rationale for tomography, see~\cite{hu_power_1999}.

\subsection{IA Correlations: References}

\begin{itemize}
    \item ``The correlation function of galaxy ellipticities produced by Gravitational Lensing"~\cite{miralda-escude_correlation_1991}\\
    \it{Historical documentation of ellipticity correlations}.

    \item ``Intrinsic alignments of galaxies in the Horizon-AGN cosmological hydrodynamical simulation"~\cite{chisari_intrinsic_2015}\\
    \it{Defines formalisms used in correlations of 3D shapes and spins}.

    \item ``Intrinsic alignments of disc and elliptical galaxies in the MassiveBlack-II and Illustris simulations"~\cite{tenneti_intrinsic_2016}\\
    \it{Defines additional formalisms used in 3D correlations}.
\end{itemize}

\section{IA Correlation Function Estimators}\label{sec:IACF}

In real or configuration space, correlations between the components described in \rsec{correlations} can be quantified using the \gls{correlation function} $\xi$.
This is often measured as a function of the separation of pairs along the \gls{LOS} ($\Pi$) and in the \gls{transverse} direction ($r_p$ for physical separation, $\bm{\vartheta}$ for angular separation on the sky).
Since we are dealing with pairs, these functions are called two-point correlation functions (2PCF).
As opposed to the power spectra \glspl{estimator} described in the next sections, these correlation functions are less sensitive to survey geometry. In practice, most observations measure the projection of $\xi$ along the LOS $w_p$. While these projected quantities contain less information than the full 3D correlations, they are more straightforward to observe and model, particularly when used in \gls{weak lensing} analyses.


\subsection{IA Correlation Function: Formalism}\label{subsec:corrfunc_form}

The IA \gls{correlation function} is most commonly measured using a generalized form of the Landy-Szalay (LS) \gls{estimator}, which was devised to estimate galaxy clustering~\citep{landy_bias_1993}. This estimator accounts for systematics and has less variance than other estimators since it uses overdensity instead of density.
To include information about alignment, the counts of galaxy pairs are weighted by the degree to which components are correlated~\citep{mandelbaum_detection_2006}.

The count of galaxy pairs between sample $A$ and $B$ is notated as $AB$.
The count weighted by the correlation between shapes in $A$ and the positions of $B$ is $A_+B$:
\begin{eqnarray}\label{eq:shape_position}
    A_+B = \sum_{i\in A, j\in B} \epsilon_+(j|i)\,.
\end{eqnarray}
The count weighted by the correlation between shapes in $A$ and shapes in $B$ is $A_+B_+$:
\begin{eqnarray}\label{eq:shape_shape}
    A_+B_+ = \sum_{i\in A, j\in B} \epsilon_+(j|i)\epsilon_+(i|j)\,.
\end{eqnarray}
Here, $\epsilon_+(j|i)$ represents the $+$ component of the \gls{ellipticity} of galaxy $j$ relative to the vector between it and the galaxy $i$ (Section~\ref{sec:Ellipticity}).
As described in \rsec{eformalism}, $0<\epsilon_+<1$ indicates \gls{radial} alignment and $-1<\epsilon_+<0$ indicates tangential alignment.

Consider a galaxy shape catalog, $S$, and a galaxy position catalog, $D$ ($D$ for density \glspl{tracer}).
These can be from the same sample. $S_+$ and $S_\times$ represent the relative $+$ and $\times$ component of shapes, as described in Eqs.~\ref{eq:shape_position},~\ref{eq:shape_shape}.
$R_D$ and $R_S$ are \gls{randoms} for the respective datasets: generated data designed to match the survey geometry but with no correlations from large-scale clustering.
The \glspl{correlation function} are
\begin{equation}
    \xi_{gg}(r_p, \Pi) = \frac{SD-R_SD-SR_D+R_SR_D}{R_SR_D}\,,
\end{equation}

\begin{equation}
    \xi_{g+}(r_p, \Pi) = \frac{S_+D - S_+R_D}{R_SR_D}\,,
\end{equation}

\begin{equation}\label{eq:intrinsic_autocorrelation}
    \xi_{++}(r_p, \Pi) = \frac{S_+S_+}{R_SR_S} \quad \text{and} \quad \xi_{\times \times}(r_p, \Pi) = \frac{S_\times S_\times }{R_SR_S}\,.
\end{equation}
Note that $\langle S_+ \rangle = \langle S_\times \rangle = 0$, which allows us to disregard the contribution of $S_+R_+$ and $S_\times R_\times$ to first order due to systematic effects.\par
For observations, the angular separation of pairs on the sky $\bm{\vartheta}$ is often used in place of $r_p$. This is done when the physical \gls{transverse} separation of galaxies cannot be easily determined, as in \gls{photometric} surveys, or when IA is being measured as a contaminant of an angular signal.

\headings{Projected Correlation Function}

Since the \gls{LOS} distance is difficult to determine in \gls{photometric} surveys and in the presence of redshift-space distortions (\gls{RSD}), most observations measure these correlation functions integrated along the LOS $\Pi$.
The projected \gls{correlation function} between two properties $a$ and $b$ is
\begin{equation}
    w_{ab}(r_p) = \int_{-\Pi_\text{max}}^{\Pi_\text{max}} \mathrm{d}\Pi\, \xi_{ab}(r_p, \Pi)\,.
\end{equation}
In practice, this requires a choice of how far along the \gls{LOS} to integrate ($\Pi_{\text{max}}$) and in what size bins ($\mathrm{d}\Pi$).
Occasionally the projected correlation function is integrated from $0-\Pi_\text{max}$ instead of $\pm\Pi_\text{max}$, in which case $\xi$ should be multiplied by 2.

\headings{3D IA correlation function}

The IA \gls{correlation function} can be measured in 3D space, such as $\xi_{g+}(r_p, \Pi)$ and $\xi_{++}(r_p, \Pi)$, or the equivalent for the $\times$ component. It can also be quantified through the monopole and quadrupole components of IA correlation as a function of the \gls{redshift}-space separation, $s$, $\xi_{g+}(s)$ or $\xi_{++}(s)$~\citep{singh_intrinsic_2016, okumura_first_2023}. While more complicated to measure and model, these \glspl{estimator} can extract significantly more information from a survey~\citep{singh_increasing_2023}.

\headings{Variable definitions}
    
\begin{itemize}
    \item $\xi$: usually used to denote a \gls{correlation function}.

    \item $\xi_{gg}$: galaxy clustering, i.e., galaxy position-position correlation.

    \item $\xi_{g+}$: cross correlation of galaxy positions and intrinsic ellipticities.

    \item $\xi_{++}, \xi_{\times\times}$: intrinsic \gls{shear} shape-shape correlation.
    
    \item $D$: positions of sample used as a density \gls{tracer}.

    \item $S$: positions of sample used for shape measurements.

    \item $R_D$: positions of random points made to match the spatial distribution of galaxy sample $D$ but with no spatial correlations.

    \item $R_S$: positions of random points  corresponding to $S$.

    \item $S_+$: shapes of galaxies in shape sample.

    \item $w_{ab}$: projected correlation function.


    \item $\Pi$: distance along the \gls{LOS}.
    $\Pi_\text{max}\sim 100 h^{-1}$Mpc.

    \item $r_p$: distance \gls{transverse} to the LOS.
\end{itemize}

\subsection{IA Correlation Function: Additional Notations}

\begin{itemize}
    \item Alternative estimators similar to the LS estimators, are, for example~\citep{joachimi_constraints_2011},
    \begin{equation}
    \xi_{g+}(r_p, \Pi) = \frac{S_+D - S_+R_D}{D_SR_D}\ \approx \frac{S_+D}{D_SR_D}\,.
    \end{equation}
    The advantages of using the LS estimator are that it is less affected by survey geometry and has a higher signal-to-noise ratio~\citep{singh_intrinsic_2015}.
    
    \item Another common way of denoting \gls{correlation function}s is to write $\xi_{\gamma g}(r_p, \Pi) = \langle\gamma(r_p) g(r_p')\rangle$.
    
    \item The projected correlation function is also notated as $w_p$.

    \item Here, we have used $\bm{\vartheta}$ to notate angular separation on the sky to distinguish it from the galaxy orientation angle. Many studies instead use $\theta$.
\end{itemize}

\subsection{IA Correlation Function: References}

\begin{itemize}
    \item ``The correlation function of galaxy ellipticities produced by gravitational lensing"~\cite{miralda-escude_correlation_1991}\\
    \it{Original formalism for combining shape information in \glspl{correlation function}.}

    \item ``The WiggleZ Dark Energy Survey: Direct constraints on blue galaxy intrinsic alignments at intermediate redshifts"~\cite{mandelbaum_wigglez_2011}\\
    \it{Definitions of the generalized LS IA estimators and helpful descriptions.}
    
    \item ``Galaxy Alignments: Observations and Impact on Cosmology"~\cite{kirk_galaxy_2015}\\
    \it{Contains a pedagogical, in-depth discussion of different IA correlation function estimators.}

    \item ``Intrinsic alignments of SDSS-III BOSS LOWZ sample galaxies"~\cite{singh_intrinsic_2015}\\
    and ``Intrinsic Alignments of BOSS LOWZ galaxies II: Impact of shape measurement methods"~\cite{singh_intrinsic_2016}\\
    \it{Contain condensed definitions of IA correlation functions with the most commonly used notation.}
\end{itemize}

\headings{Additional note: estimators derived from shear correlation functions}\label{sec:derived_shear}

Rather than using \gls{shear} \glspl{correlation function} directly, other \glspl{estimator} derived from them may be preferred, for example, to obtain higher signal-to-noise ratios or to control the physical scales used.
In \gls{weak lensing} studies, derived estimators are often chosen because they separate \glspl{$E$-mode} from \glspl{$B$-mode}.
Lensing does not produce $B$-modes so their detection in \gls{cosmic shear} data could mean that the signal is affected by systematic effects such as IA.

The references below introduce three common derived estimators: complete orthogonal sets of \textit{E/B} integrals (COSEBIs), aperture mass statistics, and band powers, and include their use for IA studies.

\begin{itemize}


    \item ``Analysis of two-point statistics of cosmic shear"~\cite{schneider_analysis_2002}\\
    \it{Illustrates the use of aperture mass statistics and band powers.}

    \item ``Sources of contamination to weak lensing three-point statistics: Constraints from N-body simulations"~\cite{semboloni_sources_2008}\\
    \it{Example of using aperture mass statistics to measure IA in simulations.}

    \item ``COSEBIs: Extracting the full \textit{E-/B}-mode information from cosmic shear correlation functions"~\cite{schneider_cosebis_2010}\\
    \it{Introduces COSEBIs as a method to separate $E$- and $B$-modes whilst retaining cosmological information.}

    \item ``KiDS-1000 cosmology: Cosmic shear constraints and comparison between two point statistics"~\cite{asgari_kids-1000_2021}\\
    \it{Discusses pros and cons of estimators.} 
\end{itemize}

\section{3D IA Power Spectrum}\label{sec:ia_ps}

The power spectrum is perhaps the most basic statistic that can be used to describe density fields.
By definition, the power spectrum is the mean square of the density fluctuation amplitudes.
It is a function of wavenumber $k$ (Fourier space) or a multipole $\ell$ (spherical harmonics space)\footnote{The choice of space will depend on many factors.
Refer to Section 3 of~\cite{kirk_galaxy_2015} for more information.}
and is the Fourier transform of the \gls{correlation function}~\citep{hikage_cosmology_2019}.

The IA power spectrum described in this section quantifies the correlation between intrinsic galaxy shapes and the galaxy (or mass) density field in Fourier space.
It is a 3D quantity that can be modeled directly or measured from simulations and is most often used to study IA directly. For a visualization of a typical IA power spectrum, see Figure 2 of~\cite{kurita_power_2021}.

\subsection{3D IA Power Spectrum: Formalism}

The Fourier transform of the intrinsic \gls{shear}, $\gamma(\bk)$, can be written as 
\begin{equation}
\begin{aligned}
    \gamma(\bk) &= \gamma_1(\bk) + i\gamma_2(\bk).
    \label{eq:def_gammaF}
\end{aligned}
\end{equation}

The galaxy-intrinsic power spectrum can be calculated from the \gls{correlation function} of the galaxy density, g, and \gls{shear}, $\gamma$,  
\begin{equation}
\begin{aligned}
    P_{g \gamma}(\bk) &= \int \int \text{d}^2 r_p \int \d \Pi\ e^{-i2\phi} e^{i\bk\cdot\mathbf{r}} \xi_{g\gamma}(r_p, \Pi)\\
    &= \int r_p \d r_p \int \d \Pi\ \xi_{g\gamma}(r_p, \Pi) \int \d \phi\ e^{-i2\phi} e^{ik r \cos \phi }\\
    &=\int r_p \d r_p \int \d \Pi\ \xi_{g\gamma}(r_p, \Pi) J_2(kr_p)\,,
    \label{eq:ps_3D}
\end{aligned}
\end{equation}
where $\vec{\bm{r}} = (r_p, \Pi)$ and $\xi_{g\gamma}(r_p, \Pi) = \langle g(r_p, \Pi) \gamma(r_p, \Pi) \rangle$~\citep{mandelbaum_wigglez_2011} and $J_2(k r_p)$ is a \gls{Bessel function} of the first kind\footnote{For more steps, see Appendix A of~\cite{ferreira_fast_2022}.}.

We can also decompose the \gls{shear} into its $E$- and $B$-modes, which are coordinate-independent quantities in Fourier space, 
\begin{equation}
    \gamma_E(\bk)+i\gamma_B(\bk) \equiv \gamma(\bk) e^{-2i\phi_{\bk}}\ ,
    \label{eq:def_EB}
\end{equation}
where $\phi_{\bk}$ is the angle measured from the first coordinate axis to $\bk_{\perp}\equiv \sqrt{(k_1^2 + k_2^2)}$, the wave vector on the sky plane, so that $\phi_{\bk} = \arctan(k_1/k_2)$.
The \gls{$E$-mode} $P_{gE}(\bk)$ and \gls{$B$-mode} $P_{gB}(\bk)$ power spectra correspond to the real and imaginary part of $P_{g \gamma}$, respectively.
 
Similarly, other types of IA power spectra are
\begin{eqnarray}
    (2\pi)^3\delta_D(\bk+\bk')P_{EE}(\bk) 
    \equiv \langle\gamma_E(\bk)\gamma_E(\bk')\rangle\,,
    \label{eq:ps_ee}\\
    (2\pi)^3\delta_D(\bk+\bk')P_{\delta E}(\bk)
    \equiv \langle\gamma_E(\bk)\delta_{\text{m}}(\bk')\rangle\,,
    \label{eq:ps_me}
\end{eqnarray}
where $\delta_{\text{m}}$ is the matter density field, and $\delta_D$ is the Dirac delta function.
Equivalently, we have,
\begin{eqnarray}
    P_{EE}(\bk) = \int \text{d}^2 r_p \d \Pi\ e^{-i\bk\cdot\mathbf{r}} \xi_{EE} (r_p,\Pi)\,, \\
    P_{E\delta}(\bk) = \int \text{d}^2 r_p \d\Pi\ e^{-i\bk\cdot\mathbf{r}} \xi_{E\delta} (r_p,\Pi)\,.
\end{eqnarray}
The \gls{$B$-mode} power spectra, $\langle \gamma_{B}\gamma_{B} \rangle = 0$ for IA caused by the scalar \gls{tidal field} in the linear regime.
$\langle g\gamma_B\rangle$ and $\langle\gamma_E \gamma_B\rangle$ should also vanish due to the statistical parity invariance of the Universe.
Note that, although \gls{$E$-mode} and \gls{$B$-mode} \gls{shear} are defined in the 2D plane perpendicular to the \gls{LOS} direction, the power spectra are functions of the 3D wave vector, $\bk$.
In other words, they are functions of $k=|\bk|$ and the direction of $\bk$.

\headings{Variable definitions}

\begin{itemize}
    \item $\bk$: 3D wave vector.


    \item $\gamma_E(\bk) = \gamma_{1}(\bk)\cos{2\phi_{\bk}} + \gamma_{2}(\bk) \sin{2\phi_{\bk}}$, \gls{$E$-mode} shear generated by the scalar gravitational potential.
    
    \item $\gamma_B(\bk) = - \gamma_{1}(\bk) \sin{2\phi_{\bk}} + \gamma_{2}(\bk) \cos{2\phi_{\bk}}$, \gls{$B$-mode} shear, that cannot be generated by the scalar field in the linear regime, used as a check for \gls{systematic errors}. These $\gamma_{E,B}$ definitions correspond to equations 9-11 in~\cite{blazek_beyond_2019}.

    \item $P_{EE}(\bk)$: auto power spectrum of the $E$-mode shear field, i.e., the intrinsic shape correlation signal.

    \item $P_{E\delta}(\bk)$: cross power spectrum between the $E$-mode shear and the mass density field, i.e., the GI signal.

    \end{itemize}

\headings{Additional note: Alternative expression for 3D IA power spectra }

By definition, any of the above power spectra can be expressed as
    \begin{equation}
    (2\pi)^3\delta_D(\bk - \bk')P{(\bk)} 
    = \langle\gamma(\bk)\gamma^*(\bk')\rangle\,,
    \end{equation}
where $^*$ is the complex conjugate,  $P{(\bk)}$ is any one of the 3D IA power spectra and $\gamma$ denotes the intrinsic \gls{shear} (or one of its components such as $\gamma_E$).

\headings{Additional note: multipole moments}

Multipole moments of the power spectrum~\citep{kurita_power_2021} can be defined as 
\begin{align}
    P_{XY}^{(\ell)}(k)=\frac{2\ell+1}{2}\int^1_{-1}\mathrm{d}\mu \mathcal{L}_\ell(\mu)P_{XY}(k,\mu), 
\end{align}
where $P_{XY}(k,\mu)$ is one of the power spectra defined in Eqs.~(\ref{eq:ps_3D}), (\ref{eq:ps_ee}) or  (\ref{eq:ps_me}), {\bm{$\mu$} is the cosine of the angle between $\bk$ and the \gls{LOS} direction,} $\mathcal{L}_\ell$ is the \gls{Legendre polynomial}.
$P^{(0)}$ is the monopole component, $P^{(2)}$ is the quadrupole component, and $P^{(4)}$ is the hexadecapole component. Additional multipoles are 0 in the large-scale (linear) regime. The angular-dependent quantities can be used to quantify information about anisotropic effects, such as \gls{RSD}.

\headings{Additional note: relation with the correlation function}

The power spectrum can be written as
    \begin{align}
    \langle[\gamma_E(\bk)+i\gamma_B(\bk)] g(\bk')\rangle
    &= \langle\gamma(\bk) g(\bk')\rangle e^{-2i\phi_{\bk}} \nonumber\\
    &\equiv (2\pi)^3 \delta_D (\bk+\bk') P_{g \gamma}(\bk)\,.
    \label{eq:ps_cross}
    \end{align}
Recall that the \gls{correlation function} is defined as
\begin{align}
    \xi_{\rm g \gamma}(\mathbf{r}) 
    = \langle\gamma(\bx) \delta_g(\bx')\rangle
    e^{-2i\phi_{\mathbf{r}}},
    \label{eq:def_2pcf_cross}
\end{align}
where $\phi_{\mathbf{r}}$ is the position angle of $\mathbf{r}\equiv \bx - \bx'$ on the sky plane.
Thus, the relation between the correlation function and the power spectrum can be written as 
    \begin{align}
    \xi_{g\gamma}(\mathbf{r}) 
    = \int \frac{\mathrm{d} \bk}{(2\pi)^3}\, 
    P_{g \gamma}(\bk) e^{2i(\phi_{\bk} - \phi_{\mathbf{r}})} e^{i\bk\cdot\mathbf{r}}.
    \end{align}
For observations, this expression can also be written as the \gls{redshift}-space \gls{correlation function}
\begin{equation}
    \xi_{AB}(r_p, \Pi, z) = \int \frac{ d^2k_\perp dk_z}{{(2\pi)^3}} P_{AB}(k, z) (1+\beta_A \mu^2)(1+\beta_B \mu^2)e^{i (r_p k_\perp + \Pi k_z)}\,,
\end{equation}
where $A,B\in\{g, +, \times\}$~\citep{singh_intrinsic_2016}. $\beta$ models the effect of \gls{RSD}~\citep{jackson_critique_1972, kaiser_clustering_1987}.
$\beta_{+}=0$ and $\beta_{\times}=0$ to first order in the case of \gls{shear}, and $\beta_g=f/b_g$ in the case of galaxies, where $f$ is the structure growth rate and $b_g$ the \gls{galaxy bias}~\citep{kaiser_spatial_1984}.

\subsection{3D IA Power Spectrum: References}

\begin{itemize}
    \item ``Power spectrum of halo intrinsic alignments in simulations"~\cite{kurita_power_2021}\\
    \it{Formalism and first measurement in N-body simulations.}
    
    \item ``Power spectrum of intrinsic alignments of galaxies in IllustrisTNG"~\cite{shi_power_2021}\\
    \it{IA power spectrum of galaxies in IllustrisTNG.}
    
    \item ``Analysis method for 3D power spectrum of projected tensor fields with fast estimator and window convolution modeling: An application to intrinsic alignments"~\cite{kurita_analysis_2022}\\
    \it{Methodology for measuring the IA power spectrum in observations.}
    
    \item ``Three-point intrinsic alignments of dark matter haloes in the IllustrisTNG simulation"~\cite{pyne_three-point_2022}\\
    \it{Extends the formalism to three-point statistics.} 
\end{itemize}

\section{2D IA Power Spectrum}\label{sec:shear_power_spectrum}

In \gls{weak lensing} analysis, observed galaxies are normally allocated to \gls{redshift} bins, so the relevant IA power spectrum is 2D\footnote{There are some exceptions where studies measure 3D \gls{cosmic shear}, e.g.~\cite{kitching_3d_2014}.}.
  It can be defined in Fourier space ($P(k)$) or in spherical harmonic space (the angular power spectrum $C(\ell)$, or simply $C_{\ell}$).
In this section, we focus on the latter.
Note that the terminology is often used quite loosely and both $P(k)$ and $C(\ell)$ may be referred to as \textit{the \gls{shear} power spectrum}.

In terms of the multipole moments of the 3D IA power spectrum, discussed in Section~\ref{sec:ia_ps}, the IA power spectrum relevant to \gls{cosmic shear} analyses is the one evaluated at $\mu=0$, i.e., $P_{\delta E}(k, \mu=0) \equiv P_{\delta E}(k_{\perp})$, where $k_{\perp} = k(1-\mu^2)^{1/2}$.

Power spectra can be expressed in terms of the corresponding \glspl{correlation function}.
In the case of IA, there is a family of correlation functions to choose from (see~\cite{kiessling_galaxy_2015} and \rsec{IACF}).
An example is the galaxy position-\gls{ellipticity} correlation function $w_{g+}(r_p)$:
\begin{equation}
    w_{g+}(r_p) = - b_g \int \d z \ W(z) \int_{0}^{\infty} \frac{\d k_{\perp} k_\perp}{2 \pi} J_2 (k_\perp r_p) P_{\delta I} (k_\perp , z)\,.
\end{equation}
Here $P_{\delta I}(k_{\perp}, z)$ corresponds to 
$P_{\delta E}(k_{\perp}, z)$ in \rsec{ia_ps}. The quantity $b_g$ is the \gls{galaxy bias}, which is assumed here to be linear, scale-independent and not to depend on \gls{redshift}. These assumptions are justifiable for quasi-linear scales where perturbation theory applies, and assuming that on these scales structure formation is entirely determined by gravity~\citep{desjacques_large-scale_2018}. At smaller scales, other considerations come into play and linear bias cannot safely be assumed~\citep{kaiser_spatial_1984, fry_biasing_1993}.

\headings{Variable definitions}

\begin{itemize}
    \item $r_p$: \gls{transverse} separation, as described in \rsec{Ellipticity}.

    \item $b_g$: \gls{galaxy bias}.

    \item $z$: \gls{redshift}.

    \item $W(z)$: redshift weighted window function~\citep{mandelbaum_wigglez_2011}. 

    \item $k_\perp$: wave vector perpendicular to the \gls{LOS}.

    \item $J_2(k_\perp r_p)$: a \gls{Bessel function} of the first kind.

    \item $P_{\delta I} (k_\perp ,z)$: represents the 3D density-ellipticity power spectrum.
\end{itemize}

\begin{figure*}[h]
\centering
\includegraphics[width=0.5\textwidth]{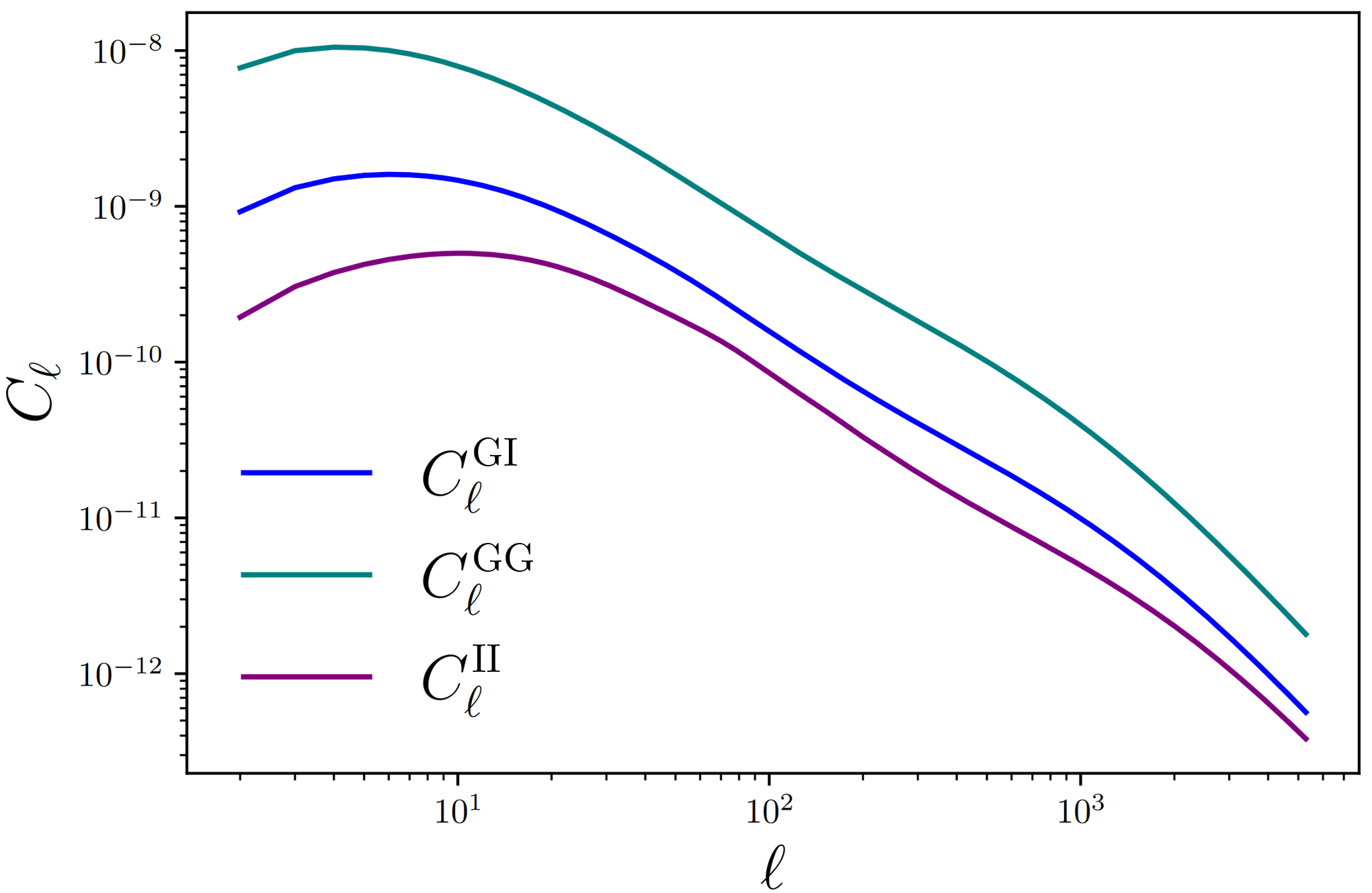}
\caption[Examples of angular power spectra for the different contributions given in \req{2D_angular_spectrum}. Note that these are highly dependent upon \gls{redshift} distribution and modeling. The $C_\ell^{\text{GI}}$ curve shows the absolute value for the component, as it is an anti-correlation. This plot was generated using the python package \texttt{pyccl}~\citep{chisari_cosmology_2019}, assuming the nonlinear alignment model described in \rssec{NLA}. \vspace{.2in}]{\renewcommand{\thempfn}{\arabic{mpfootnote}}
Angular power spectra for the different contributions given in \req{2D_angular_spectrum}. The $C_\ell^{\text{GI}}$ curve shows the absolute value for the component, as it is an anti-correlation. This plot was generated using the python package \texttt{pyccl}\protect\footnotemark[4]~\citep{chisari_cosmological_2013}, assuming the nonlinear alignment model described in \rssec{NLA}. \vspace{.2in}}
\label{fig:cls}
\end{figure*}

The 3D IA power spectrum can be projected along the \gls{LOS} to construct the \textbf{2D projected angular IA power spectrum} $C_{\epsilon \epsilon}^{ij} (\ell)$ as the sum of the gravitational lensing part GG, gravitational-intrinsic GI, and intrinsic-intrinsic contribution II:

\begin{equation}\label{eq:2D_angular_spectrum}
    C _{\epsilon \epsilon} ^{ij} (\ell) = C _{\mathrm{GG}} ^{ij} (\ell) + C ^{ij} _{\mathrm{GI}} (\ell)  + C _{\mathrm{II}} ^{ij} (\ell)\,.
\end{equation}
Each contribution in the above expression is described in spherical harmonic space where $\ell$ is the angular frequency. \rfig{cls} shows the $C_{\ell}s$ for each of these components.

Indices $(i, j)$ represent two redshift bins in which the correlation is taking place (see Section~\ref{note:tomography}).
Terms in \req{2D_angular_spectrum} can be represented as:
\begin{align}
    C _{\mathrm{GG}}^{ij} (\ell) = \int_{0}^{\chi_{\mathrm{H}}} \d \chi \frac{q^i (\chi) q ^j (\chi)}{f^2_K (\chi)} P_{\delta \delta} \left( k = \frac{\ell}{f_K (\chi)} , \chi \right)\,, \\
    C _{\mathrm{GI}}^{ij} (\ell) = \int_{0}^{\chi_{\mathrm{H}}} \d \chi \frac{p^i (\chi) q ^j (\chi)}{f^2_K (\chi)} P_{\delta \mathrm{I}} \left( k = \frac{\ell}{f_K (\chi)} , \chi \right)\,, \\
    C _{\mathrm{II}}^{ij} (\ell) = \int_{0}^{\chi_{\mathrm{H}}} \d \chi \frac{p^i (\chi) p ^j (\chi)}{f^2_K (\chi)} P_{\mathrm{II}} \left( k = \frac{\ell}{f_K (\chi)} , \chi \right)\,.
\end{align}
\rsec{Modeling} discusses various methods used to model the power spectra $P_{\delta \mathrm{I}}$ and $P_{\mathrm{II}}$.
\footnotetext[4]{\href{https://github.com/LSSTDESC/CCL}{https://github.com/LSSTDESC/CCL}}

\headings{Variable definitions}

\begin{itemize}
    \item $\chi$ is the \gls{radial} comoving distance.

    \item $\chi_\mathrm{H}$ is the comoving distance to the horizon.

    \item $f_K (\chi)$ is the radial function:
    \[
    f_K (\chi) =
    \begin{cases}
    \frac{1}{\sqrt{K}} \sin{(\sqrt{K} \chi)},
    & 
    \text{if } 
    \;  \; K > 0 
    \;  \; \text{(open universe)}\\
    
    \chi, 
    & \text{if } 
    \;  \; K = 0 
    \;  \;  \text{(flat universe)}\\
    
    \frac{1}{\sqrt{-K}} \sinh{(\sqrt{-K} \chi)}, & \text{if} 
    \;  \; K < 0 
    \;  \; \text{(closed universe)}
    
    \end{cases}
    \]
    with $\frac{1}{\sqrt{|K|}}$ the curvature radius of the spatial part of spacetime.

   \item $q^i (\chi)$ and $q^j (\chi)$ are the lensing weighting functions in tomographic bins $i$ and $j$, respectively: 
   \begin{equation}
        q^i (\chi) = \frac{3 H_0^2 \Omega_{\text{m}}}{2 c^2} \frac{f_K (\chi)}{a (\chi)} \int_{\chi}^{\chi_{\mathrm{H}}} \d \chi' p^i (\chi') \frac{f_K (\chi' - \chi)}{f_K (\chi')}\,,
   \end{equation}
   with $H_0$ the Hubble parameter, $\Omega_{\text{m}}$ the matter density fraction, $c$ the speed of light in a vacuum, and $a(\chi)$ the scale factor.

    \item $W$: The window function is often called a filter function or a survey window.
    As well as \gls{redshift} $z$, it may be defined in terms of other survey parameters: position on the sky $\theta$, etc.
    \item $p^i (\chi)$ and $p^j (\chi)$ are the redshift distributions in tomographic bins $i$ and $j$, respectively, in radial comoving distance space:
    $p^i (\chi) = p^i (z) \frac{\d z}{\d \chi}$ and $p^j (\chi) = p^j (z) \frac{\d z}{\d \chi}$.
\end{itemize}

\headings{Limber approximation}\label{note:limber_approximation}

Power spectra on small scales (high multipoles $\ell$) are computationally expensive to calculate, especially if they involve rapidly oscillating functions.
In practical applications, the Limber approximation  is often used~\citep{loverde_extended_2008,kitching_limits_2017}.
In this approximation, the integrand varies more slowly but the result is still accurate at small scales.

One way to understand the approximation is to write the lensing angular power spectrum, $C_\ell^{\phi\phi}$, as~\citep{lemos_effect_2017}
\begin{equation}
 C_\ell^{\phi\phi}(r,s)= \frac{8}{\pi}\left(\frac{3 \Omega_{\text{m}}^2 H_0^2 }{2 c^2} \right)^2  \int\frac{\d k}{k^2} \ I^r_\ell(k)I^s_\ell(k)\,,
\end{equation}
where
\begin{equation}
    I^r_\ell(k) = \int \frac{\mathrm{d}\chi}{\chi} \ [1+z(\chi)]q^r(\chi)j_\ell(k\chi)[P_\delta(k;\chi)]^{1/2}\,.
\end{equation}
The Limber approximation replaces the spherical \gls{Bessel function} $j_\ell(k\chi)$ with a delta function:
\begin{equation}
    j_\ell (k\chi) \rightarrow \sqrt{\frac{\pi}{2 \nu}} \delta_D (\nu - k_\chi)\,,
\end{equation}
with $\nu = \ell + \frac{1}{2} = k\chi$.
Then the \gls{shear} power spectrum takes the form:

\begin{equation}
    C_\ell ^{\epsilon \epsilon} (r, s) = \frac{(\ell + 2)!}{\nu^4 (\ell - 2)!} \left( \frac{3 \Omega_{\text{m}}^2 H_0^2 }{2 c^2} \right)^2 \int \d \chi \ [1 + z (\chi)]^2 q^r (\chi) q^s (\chi) P_\delta \left( \frac{\nu}{\chi} ; \chi \right)\,.
\end{equation}
The Limber approximation assumes that the variation of the kernels of the projected fluctuations is limited to scales larger than the average clustering length, which makes it invalid across all scales. 
This limitation, combined with the need to reduce \gls{systematic errors} for future cosmological surveys, requires non-Limber methods.
There have been recent efforts to move away from the Limber approximation in numerical analyses, such as in~\cite{abbott_dark_2023} and~\cite{leonard_n5k_2023}, which presents several alternatives to implement this calculation in a fast and reliable framework.

\subsection{2D IA Power Spectra: Additional Notations}

\begin{itemize}
    \item $r$ and $s$: indices that denote two tomographic bins are usually $i$ and $j$.
    Some authors use $r$ and $s$ instead (for example~\cite{lemos_effect_2017}).
   
    \item $n^r (\chi)$ and $n^s (\chi)$: often used in place of $p^i (\chi)$ and $p^j (\chi)$ to denote the \gls{redshift} distributions in respective tomographic bins $i$ and $j$.
\end{itemize}

\subsection{2D IA Power Spectra: References}

\begin{itemize}
    \item ``Weak lensing power spectra for precision cosmology: Multiple-deflection, reduced \gls{shear} and lensing bias corrections"~\cite{krause_weak_2010}\\
    \it{Detailed derivation for the corrected \gls{weak lensing} power spectra.}

    \item ``Cosmology from cosmic shear power spectra with Subaru Hyper Suprime-Cam first-year data"~\cite{hikage_cosmology_2019}\\
    \it{Example of accounting for IA in power spectrum analysis of data from the Hyper Supreme-Cam (HSC).}


    \item ``KiDS-1000 cosmology: Cosmic shear constraints and comparison between two point statistics"~\cite{asgari_kids-1000_2021}\\
    \it{Short breakdown of the power spectra for the KiDS survey, containing a list of helpful references.}
\end{itemize}

\headings{Additional note: beyond the power spectrum}

Power spectra are powerful probes of IA.
However, 2-point statistics, such as the power spectrum or the 2PCF, fully characterize all the available information only for Gaussian and isotropic fields. Ideally, we want to ensure that high-order statistics (e.g., 3-point correlations \citep{schmitz_time_2018,pyne_three-point_2022}) are taken into account when inferring the IA signal, and that uncertainties are accounted for self-consistently via \gls{forward modeling} (e.g., via field-level inferences \citep{tsaprazi_field-level_2022, porqueres_field-level_2023}). With the advent of next-generation data, these methods must be advanced to smaller scales in order to increase the signal-to-noise ratio of the detections.

\section{Modeling}\label{sec:Modeling}

Many IA models have been developed over the past two decades and even the earlier ones are still useful.
Advancements reflect the increasing understanding and importance of IA and the need to extend theory to smaller scales and a wider range of alignment mechanisms.
Models are often created for specific scales and galaxy types, the main two being \gls{spiral} and \gls{elliptical}.
As noted in Section~\ref{note:galaxy_types}, how these morphologies relate to \gls{ELG}/\glspl{LRG} is a source of discussion and the distinction between ``red'' and ``blue'' galaxy samples is survey- and simulation-dependent.

This section briefly summarizes common models with the aim of introducing typical modeling formalisms; it is not a complete overview of the field.
In Sections~\ref{subsec:lin_alignment_model} and \ref{subsec:NLA} we discuss the linear and nonlinear alignment models. These were developed first and are the most frequently used to date. Section~\ref{subsec:ia_amp} expands on aspects of the intrinsic alignment amplitude parameter which is a feature of these models.
Sections~\ref{subsec:TATT} to \ref{subsec:halomodel} introduce three more well-established approaches to modeling IA: the Tidal Alignment and Tidal Torquing model, effective field theory, and the \gls{halo} model. Section~\ref{subsec:other_models} briefly highlights some more recent models. \rssec{obs_status} summarizes the observational status of the models. Finally, \rssec{self_cal} introduces self-calibration, a model-agnostic technique.

\subsection{Linear Alignment model (LA)}\label{subsec:lin_alignment_model}

This approach assumes that the \gls{ellipticity} of a galaxy is linearly related to the gravitational potential at the time the galaxy formed.
Before we start, it is useful to define two \glspl{redshift} relevant for the LA and other models:
\begin{itemize}
    \item \textbf{$z_{\mathrm{IA}}$}: the redshift at which the alignment is set.

    \item \textbf{$z_\mathrm{obs}$}: the redshift at which the IA is observed.
\end{itemize}

Since the details of galaxy formation and evolution are not well understood, two scenarios are proposed for the \gls{redshift} $z_\text{IA}$ at which the alignment is produced.

\begin{itemize}
    \item \textbf{Instantaneous Alignment:} In this case the alignment redshift is assumed to be the same as the observed \gls{redshift}, $z_{\mathrm{IA}} = z_{\mathrm{obs}}$, and the amplitude of the alignment, $A_\mathrm{IA}$, is modeled as a simple linear function of redshift.
    \begin{equation}
    A_\mathrm{IA} (z) = - C_1 \rho_{\text{m},0} (1 + z)\,,
    \end{equation}
    where $C_1$ is a normalization constant (a contribution due to the response to the \gls{tidal field}) and $\rho_{\text{m},0}$ is the matter density at $z = 0$ (today).

    \item \textbf{Early Alignment:}
    In the early alignment scenario, the \gls{redshift} at which the IA signal is observed is later than the redshift at which the alignment was set,  $z_\mathrm{obs} <z_\mathrm{IA}$.
    If it is assumed that the \gls{halo} (or galaxy) first forms during the matter domination era (primordial alignment) then $A_\mathrm{IA}$ does not depend on $z_\mathrm{IA}$ and evolves as
    \begin{equation}
    A_\mathrm{IA} (z) = - \frac{C_1 \rho_{\text{m}, 0} (1 + z)}{\bar{D} (z)}\,, 
    \end{equation}
    where $\Bar{D} (z)$ has been set to unity at matter domination: $\Bar{D} (z) = (1 + z) D (z)$, and $D(z)$ is the growth factor.
    Alternatively, it is often assumed that alignment happens later than this because it is affected by more recent mergers and accretion.
    Then the amplitude is given by:
    \begin{equation}
    A_\mathrm{IA} (z) = - C_1  \rho_{\text{m},0} (1 + z_{\mathrm{IA}}) \frac{D (z_\mathrm{IA})}{D (z)}\,.
    \end{equation}
    Finally, in each of these cases, the IA power spectra are: 
    \begin{align}
        P_{\delta \mathrm{I}}(k,z) &= A_\mathrm{IA}(z) P_\mathrm{L}(k,z) \label{eq:LA1} \\
        P_\mathrm{II}(k,z) &= A^2_\mathrm{IA}(z) P_\mathrm{L}(k,z)\,,\label{eq:LA2}
    \end{align}
    where $P_\mathrm{L}(k,z)$ is the linear matter power spectrum.
\end{itemize}

\subsubsection{LA model: References}

\begin{itemize}
    \item ``Intrinsic and extrinsic galaxy alignment"~\cite{catelan_intrinsic_2001}\\
    \it{Introduction of the LA model.}

    \item ``Intrinsic alignment-lensing interference as a contaminant of cosmic shear"~\cite{hirata_intrinsic_2004}\\
    \it{Development of the LA model.}

    \item{``Tidal alignment of galaxies"}~\cite{blazek_tidal_2015}\\
    \it{Section 3.2 discusses alignment epochs.}
\end{itemize}

\subsection{Nonlinear Alignment model (NLA)}\label{subsec:NLA}

This is an empirical model that replaces the linear power spectrum used in the LA model  (Eqs. \ref{eq:LA1} and \ref{eq:LA2}) with the full nonlinear matter power spectrum, $P_\text{NL}(k,z)$, while preserving the assumption that density perturbations are described by the Poisson equation.
Physically, the latter assumption fails in the nonlinear regime of structure formation due to gravitational evolution.
However, this model has been widely adopted due to its ability to accurately reproduce \gls{ellipticity} correlations for red \gls{elliptical} galaxies down to $\sim6$ $h^{-1}$Mpc.
The NLA model is often enhanced to include other sample dependencies, for example, on \gls{redshift} or galaxy luminosity, usually captured by additional power laws.

To incorporate a luminosity $L$ dependence, the IA amplitude $A_\text{IA}$ is expressed as a power law of the general form:
\begin{equation}
A_{\text{IA}} \mapsto A \left( \frac{L}{L_0}\right)^{\alpha_L}\,,
\end{equation}
where $A$ is a prefactor, $L_0$ is a \gls{pivot luminosity} 
and $\alpha_L$ is a luminosity scaling.
Furthermore, the IA amplitude form can be extended to account for the \gls{redshift} evolution as
\begin{equation}
    A_{\text{IA}} (L, z) = A \left( \frac{L}{L_0} \right)^{\alpha_L} (1 + z)^{\alpha_z}\,,
\end{equation}
with $\alpha_z$ a redshift scaling parameter.
This parameter is often denoted $\eta$ or, if the redshift contribution is separated into low- and high-redshift parts, $\eta_{\mathrm{low-z}}$ and $\eta_{\mathrm{high-z}}$.
This should not be confused with the definition in \rsec{ellipticity_addnotation} or in \rsec{correlations}.
For more details, consult~\cite{chisari_redshift_2016}.

\headings{Practical uses}

Another parameterization of the \gls{redshift}- and luminosity-dependent IA amplitude frequently used in analyses is:
\begin{equation}
    A_\mathrm{IA} (L, z) = A_0 \frac{C_1 \rho_{\text{m},0}}{D (z)} \left( \frac{L}{L_0} \right)^{\alpha_L} \left( \frac{1 + z}{1 + z_0} \right)^{\alpha_z}\,,
\end{equation}
where $A_0$ is a prefactor and $z_0$ is a pivot redshift.
$C_1$, $\rho_{\text{m},0}$ and $D (z)$ are described in Section~\ref{subsec:lin_alignment_model}.
Since the signal is not observed in blue galaxies, the amplitude is corrected for the fraction of red galaxies and the redshift scaling can be separated into ``low redshift" and ``high redshift" contributions.
The details can be found in~\cite{krause_impact_2016}.

\subsubsection{NLA model: References}

\begin{itemize}
    \item ``Dark energy constraints from cosmic shear power spectra: Impact of intrinsic alignments on photometric redshift requirements"~\cite{bridle_dark_2007}\\
    \it{First use of the NLA model.
    The authors argue that this model might not be ``closer to the truth", but that it matches the data slightly better than the LA model.}

    \item ``Constraints on intrinsic alignment contamination of weak lensing surveys using the MegaZ-LRG sample"~\cite{joachimi_constraints_2011}\\
    \it{Frequently cited paper giving commonly used value for the constant $C_1$ in the LA/NLA models (though the original source of the value can be traced back via~\cite{bridle_dark_2007},~\cite{hirata_intrinsic_2004} and~\cite{brown_measurement_2002})}.

    \item{``Intrinsic alignments of SDSS-III BOSS LOWZ sample galaxies"}~\cite{singh_intrinsic_2015}\\
    \it{Tests the NLA model against observations.}

    \item ``The impact of intrinsic alignment on current and future cosmic shear surveys"~\cite{krause_impact_2016}\\
    \it{Detailed description of the luminosity- and \gls{redshift}-dependent IA amplitude.}

\end{itemize}

\subsection{Alignment Amplitude}\label{subsec:ia_amp}

The alignment amplitude $A_{\text{IA}}$ in these models is essentially a free bias parameter that relates a measurement of the local \gls{tidal field} to the amplitude, or strength, of the IA signal. 
When cosmological parameters are estimated from \gls{weak lensing}, $A_{\text{IA}}$ is often treated as a ``nuisance" parameter to be marginalized over.

As discussed in Sections \ref{subsec:lin_alignment_model}  and \ref{subsec:NLA}, the amplitude is expected to evolve with \gls{redshift} and also to depend on galaxy/\gls{halo} mass.
The redshift dependence is thought to arise because galaxies traverse evolving regions of tidal \gls{shear} (advection)~\citep{schmitz_time_2018} or because halo mass accretion is stronger at higher redshifts~\citep{asgari_halo_2023}. 

In simulations $A_{\text{IA}}$ has been found to depend on several galaxy properties including mass, luminosity (increasing with both), color, and morphology (stronger for red/\gls{elliptical} than blue/\gls{spiral}) \citep{samuroff_advances_2021}. Observations have found $A_{\text{IA}}$ to be consistent with zero for blue galaxies~\citep{johnston_kidsgama_2019}, suggesting a lack of \gls{tidal alignment}, but robust detections have been made for red galaxies, e.g.,~\cite{fortuna_kids-1000_2021}.

All these dependencies are difficult to quantify in practice.
Because of this, the \gls{redshift} scaling is often separated into ``low redshift" and ``high redshift" contributions.
Alternatively, the alignment amplitude is simply set to be constant across the redshift range.
The amplitude may also be corrected for the fraction of red galaxies, or the galaxies can be divided into ``red" and ``blue" sub-samples which are analyzed separately (for example~\cite{samuroff_dark_2019}).
As discussed in \rssec{NLA}, in observational studies luminosity is often used as a proxy for mass so that we have $A_{\text{IA}}(z, L)$.
A further complication is that the observed alignment amplitude has been found to depend on the method used to measure galaxy shapes {\citep{singh_intrinsic_2016}}.

\subsubsection{IA Amplitude: Additional Notation}

Although $A_{\text{IA}}$ is a common notation, the amplitude is sometimes denoted as $A$ or $A_{\text{I}}$.
However, one must be careful as they often represent different quantities depending on the model.
Additionally, superscripts may be used to indicate the amplitude of subsamples e.g., $A_\text{IA}^\text{R}$, $A_\text{IA}^\text{B}$ for red and blue samples.

For models with more than one IA amplitude the notation $A_1$, $A_2, \ldots$ is often used.
In this case, $A_1$ may also be used for the single-amplitude model.

\subsubsection{IA Amplitude: References}

\begin{itemize}
    \item ``Redshift and luminosity evolution of the intrinsic alignments of galaxies in Horizon-AGN"~\cite{chisari_redshift_2016}\\
    \it{A detailed work on the IA amplitude dependency on \gls{redshift} and luminosity.}

    \item ``KiDS+GAMA: Intrinsic alignment model constraints for current and future weak lensing cosmology"~\cite{johnston_kidsgama_2019}\\
    \it{Dependency of IA amplitude on color in observations.}

    \item ``KiDS-1000: Constraints on the intrinsic alignment of luminous red galaxies"~\cite{fortuna_kids-1000_2021}\\
    \it{Luminosity and redshift dependence of IA for \glspl{LRG}.}

    \item ``Advances in Constraining Intrinsic Alignment Models with Hydrodynamic Simulations"~\cite{samuroff_advances_2021}\\
    \it{Comprehensive analysis comparing multiple simulations and dependence of IA on galaxy properties.}
\end{itemize}

\subsection{Tidal Alignment and Tidal Torquing model (TATT)}\label{subsec:TATT}

The LA and NLA models are strictly thought to apply only to \gls{elliptical} galaxies.
TATT was introduced to account for the alignment of \gls{spiral} galaxies whose configuration depends on angular momentum rather than being pressure-supported.
The basic idea is to express a galaxy's intrinsic shape $\gamma^\mathrm{I}_{ij}$ as an expansion of the \gls{trace-free} \gls{tidal field} tensor $s_{ij}$ to second order in the linear density field:
\begin{equation}
   \gamma^\mathrm{I}_{ij}= \bar{C}_1  s_{ij}+ \bar{C}_{1\delta}(\delta \times s_{ij}) + \bar{C}_2 \bigg[\sum_{k=0}^2 s_{ik}s_{kj} - \frac{1}{3}\delta_{ij}s^2\bigg]\,,\label{eq:TATT}
\end{equation}
where $\delta$ and $s$ respectively describe the density and tidal fields. Importantly, the TATT model does not include all possible terms of the perturbative bias expansion to second order \citep{blazek_beyond_2019}.

The first term of Eq.~\ref{eq:TATT} corresponds to \gls{tidal alignment} as in the LA/NLA models.
The second is ``density weighting" and arises because alignment can only be measured where there is a galaxy.
The third term is tidal torquing.

Tidal alignment is described by 
\begin{equation}
\bar{C}_1(z) = - A_1(z) \frac{C_1\rho_{\text{m}}}{D(z)}\,,
\end{equation}
where $C_1$ is the same normalization constant as in the LA and NLA models.
A similar formula applies to $\bar{C}_{1\delta}$, which can be parameterized as $\bar{C}_{1\delta}= b_1 \bar{C}_1$, where $b_1$ is called the linear bias.
$A_1$ is equivalent to $A_\mathrm{IA}$ in the LA/NLA models.
As for the LA model (Section \ref{subsec:lin_alignment_model}), adjustments can be made to allow for different assumptions about the redshift at which alignment is set. 

The tidal torquing parameter, $\bar{C}_2$, is given by
\begin{equation}
  \bar{C}_2(z) = A_2(z) \frac{5C_1\rho_{\text{m}}}{D^2(z)}\,,  
\end{equation}
where $A_2$ is a second alignment amplitude.
The factor 5 is an approximation related to the different IA power spectrum in the pure \gls{tidal alignment} and tidal torquing cases.

\subsubsection{TATT model: References}

\begin{itemize}
    \item ``Beyond linear galaxy alignments"~\cite{blazek_beyond_2019}\\
    \it{Introduction of the TATT model and full equations for IA power spectra}.

    \item ``Advances in constraining intrinsic alignment models with hydrodynamic simulations"~\cite{samuroff_advances_2021}\\
    \it{2PCF measurements from simulations comparing the NLA and TATT models (and comparing simulations).
    This contains a helpful discussion of the models.}
\end{itemize}

\subsection{Effective Field Theory (EFT) model}\label{subsec:EFT}

EFT extends the modeling of IA to smaller scales in a more theoretically rigorous way than the NLA model. It is similar to Eulerian perturbation theory which has previously been applied to scalar fields such as counts of galaxies (for example~\cite{baumann_cosmological_2012}). In the case of IA we instead have a tensor field, but in other respects the method is similar. 
On large scales, the matter density distribution is treated as an effective fluid, while the small-scale physics are decoupled and packed into a set of free parameters with values that can be constrained through either simulations or observations.

For IA, the EFT approach involves first capturing the physical interactions that give rise to intrinsic galaxy shape correlations within the galaxies' rest frame.
These correlations are then expressed in terms of local gravitational observables.
This accounts for the statistical properties of galaxy shapes resulting from gravitational effects and allows for a systematic approach to modeling the corresponding IA.

The shape of a galaxy can be defined by a \gls{trace-free} symmetric tensor $g_{ij}$ based on its light distribution.
The shape tensor is then connected to local gravitational observables by a bias expansion
\begin{equation}
    g_{ij} = \sum_O  b_O^{(g)}[O_{ij}]\,,
\end{equation}
where $[O]$ are renormalized operators and $b_O$ are corresponding bias parameters. This is exactly the same approach as for a scalar field.  The next step is to determine which operators should be included to any given order in perturbation theory.  
Since these include higher spatial derivatives of the density and \glspl{tidal field}, the modeling of IA is extended from linear scales to the quasi-linear regime.

In the case of IA, the final step is to project the obtained statistical information onto the sky, effectively mapping the correlations onto the observed galaxy shapes~\citep{vlah_eft_2020}. The EFT  model can characterize the auto-spectrum of IA \glspl{$B$-mode} with high accuracy~\citep{bakx_effective_2023} and has been used to predict the $B$-mode IA of clusters~\citep{georgiou_b-modes_2023}.


\subsubsection{EFT: References}

\begin{itemize}
    \item ``Cosmological Non-Linearities as an Effective Field"~\cite{baumann_cosmological_2012}\\
    \it{Example of use of EFT of large-scale structure.}
    
     \item ``Large-Scale Galaxy Bias"~\cite{desjacques_large-scale_2018}\\
    \it{Review article. Appendix B.3 discusses the EFT of large-scale structures.}
    
    \item ``An EFT description of galaxy intrinsic alignments"~\cite{vlah_eft_2020}\\
    \it{EFT description of 3D IA.}

    \item ``Galaxy shape statistics in the effective field theory"~\cite{vlah_galaxy_2021}\\
    \it{Projection onto the observed sky.}

    \item ``Effective field theory of intrinsic alignments at one loop order: a comparison to dark matter simulations"~\cite{bakx_effective_2023}\\
    \it{Comparison of EFT of IA to simulations.}
\end{itemize}


\subsection{Halo model}\label{subsec:halomodel}

The \gls{halo} model of IA \citep{schneider_halo_2010} is an alternative approach that builds on the halo model of galaxy clustering \citep{cooray_halo_2002}. In this model, all matter is assumed to reside in dark matter halos and the galaxy power spectrum is expressed as the sum of a one-halo term corresponding to the (small-scale) correlation of galaxies within a single dark matter halo, and a two-halo term corresponding to the (large-scale) correlation of galaxies in different halos.

The same approach can be taken for the IA power spectra so that
\begin{align}
P_\mathrm{GI}(k) &= P^\mathrm{1h}_\mathrm{GI}(k)+P^\mathrm{2h}_\mathrm{GI}(k) \\
P_\mathrm{II}(k) &= P^\mathrm{1h}_\mathrm{II}(k)+P^\mathrm{2h}_\mathrm{II}(k)\,.
\end{align}
From now on we consider only $P_\mathrm{GI}(k)$. In all cases equivalent expressions can be derived for $P_\mathrm{II}(k)$. 

Crucially, this formalism enables different modeling of the alignment of \gls{satellite} and \gls{central} galaxies, which are thought to contribute to the IA power spectra in different ways and on different scales. 
It can be assumed that centrals contribute mainly to the two-halo term, that they have the same \gls{ellipticity} and orientation as their parent halos, and that this large-scale behavior can be described by the linear alignment model.
Moreover, the two-halo term can easily be split to allow different amplitudes for the alignment contribution of red and blue populations:
\begin{equation}
P^\mathrm{2h}_\mathrm{GI}(k) = f^\mathrm{red}_\mathrm{c} P^\mathrm{red}_\mathrm{GI}(k) + f^\mathrm{blue}_\mathrm{c} P^\mathrm{blue}_\mathrm{GI}(k)\,,
\end{equation}
where the fractions of red and blue centrals, $f^\mathrm{red}_\mathrm{c}$ and $f^\mathrm{blue}_\mathrm{c}$, can be determined using a model of the \gls{halo occupation distribution}.

The alignment of \glspl{satellite} has been found to be weaker and more complex. At small scales, satellites are often assumed to point radially toward the center of their host halo. At large scales there is evidence that their alignment vanishes, suppressing the overall signal. The halo model can take this complexity into account.

Under the assumption of a spherical halo, the inter-halo alignment between \glspl{central} and \glspl{satellite} is zero on average, and thus the one-halo term describes the alignment of satellites with each other and the local matter field:
\begin{equation}
P^\mathrm{1h}_\mathrm{GI}(k) = \int \d M n(M) \frac{M}{\bar{\rho}_{\text{m}}} f_\mathrm{s} \frac{\langle N_\mathrm{s}|M \rangle}{\bar{n}_\mathrm{s}} |\hat{\gamma}^\mathrm{I}(\boldsymbol{k}|M)| \hat{U}(M,k)\,,
\end{equation}
where $n(M)$ is the \gls{halo mass function}, $f_\mathrm{s}$ is the fraction of satellite galaxies, which can be written as a function of \gls{redshift}, $\langle N_\mathrm{s}|M \rangle$ is the \gls{halo occupation distribution} of satellites, $\hat{U}(M,k)$ is the normalized mass density profile, and $\hat{\gamma}^\mathrm{I}$ is the density-weighted average of the satellite intrinsic ellipticity.
An additional dependence on radius can also be included.

The halo model has been shown to fit observations in the range $0.3 - 1.5$ $h^{-1}$Mpc~\citep{singh_intrinsic_2015}, and on larger scales can be complemented by the LA or the NLA models.
Although its formalism relies on physical assumptions, such as the symmetry of the halos or the alignment of \glspl{satellite} inside a halo, and it suffers from the lack of constraints for fainter galaxies, this model performs well in the context of \gls{weak lensing} corrections for current surveys.

\headings{Variable definitions}

\begin{itemize}
    \item $P^\mathrm{2h}_\mathrm{GI}(k)$: two-halo term of the gravitational-intrinsic power spectrum.

    \item $f^\mathrm{red}_\mathrm{c},f^\mathrm{blue}_\mathrm{c}$: fractions of red and blue \gls{central} galaxies.

    \item $P^\mathrm{1h}_\mathrm{GI}(k)$: one-halo term of the gravitational-intrinsic  power spectrum.

    \item $n(M)$: \gls{halo mass function}.

    \item $f_\mathrm{s}$: fraction of \gls{satellite} galaxies.

    \item $\langle N_\mathrm{s}|M \rangle$: \gls{halo occupation distribution} of satellites.

    \item ${\bar{n}_\mathrm{s}}$: mean number density of galaxies.

    \item $|\hat{\gamma}^\mathrm{I}(\boldsymbol{k}|M)|$: density-weighted average satellite ellipticity.

    \item $\hat{U}(M,k)$: normalized mass density profile.
\end{itemize}

\subsubsection{Halo model: References}

\begin{itemize}
    \item ``Halo models of large scale structure"~\cite{cooray_halo_2002}\\
    \it{The original halo model of galaxy clustering.}

    \item ``A halo model for intrinsic alignments of galaxy ellipticities"~\cite{schneider_halo_2010}\\
    \it{The first IA halo model.}

    \item ``The halo model as a versatile tool to predict intrinsic alignments"~\cite{fortuna_halo_2021}\\
    \it{Development of a halo model for IA.}

    \item ``The halo model for cosmology: A pedagogical review"~\cite{asgari_halo_2023}\\
    \it{Introduction to the halo model with some discussion of IA.}
\end{itemize}

\subsection{Other IA  models}\label{subsec:other_models}
Additional modeling approaches have been proposed to address weaknesses in existing models and to take advantage of increased computational resources. This is a rapidly-moving field and so here we only mention three examples.

\begin{itemize}
    \item{\textbf{Semi-analytic models}: These assign shapes and orientations to galaxies placed in dark-matter-only simulations.  The approach is not new, but a recent example using up-to-date simulations is~\cite{hoffmann_modeling_2022}}.
     \item {\textbf{HYMALAIA}~\citep{maion_hymalaia_2023}: This is a hybrid Lagrangian model that combines a perturbative bias expansion in Lagrangian space with the fully nonlinear displacement field from N-body simulations. It has been found to be more accurate than TATT and to be valid up to smaller scales while having fewer free parameters than EFT models.}     
    \item  {\textbf{Lagrangian perturbation theory}~\citep{chen_lagrangian_2024}:
    Similar to the Eulerian EFT model in~\cite{vlah_eft_2020} but with a different treatment of long-wavelength linear displacements and an alternative formulation of the bias scheme.}
    
\end{itemize}

\subsection{Modeling: Observational status}\label{subsec:obs_status}

The assumption of the linearity of tidal \gls{shear} breaks down on small scales, where linear models under-predict correlations.
Therefore, alignment models that are more informed about nonlinear structure growth are required for accurate modeling of galaxy shapes.
\cite{chisari_intrinsic_2014} reported that the linear alignment model behaves well only down to $10$ $h^{-1}$Mpc, and~\cite{singh_intrinsic_2015} found that the nonlinear alignment model accurately models the ellipticities of luminous red galaxies down to $6$ $h^{-1}$Mpc.
The TATT model attempts to extend scales down to $2$ $h^{-1}$Mpc~\citep{samuroff_dark_2022}, though does not describe the alignment of \glspl{halo} at quasi non-linear regimes as the full EFT model does \citep{bakx_effective_2023}.
Finally, the halo model has been successfully used in the range $0.3 - 1.5$ $h^{-1}$Mpc~\citep{singh_intrinsic_2015}, yet also lacks the ability to fully capture correlations in the nonlinear regime.

\subsection{Self-calibration}\label{subsec:self_cal}

When \gls{weak lensing} data is used to constrain cosmological parameters, an alternative to directly modeling IA and marginalizing over parameters is to follow a model-agnostic approach through self-calibration.

The approach is based on the insight that in \gls{photometric} surveys it is possible to identify features of the GG (lensing) power spectrum that distinguish it from the GI and II spectra. \cite{zhang_proposal_2010} 
considered the dependence of the power spectra on the difference $\Delta z^\text{P}$ between the photometric \glspl{redshift} of two tomographic bins and showed that the relationship for GG can be expected to be quite different from that for GI or II. This is essentially because of the different weighting functions in the expressions for the power spectra (see Section~\ref{sec:shear_power_spectrum}). It is then possible to derive scaling relations for the IA power spectra in terms of $\Delta z^\text{P}$ and non-IA observables. Thus the IA elements of the observed power spectrum can be removed without making any assumptions about the form of the IA and without needing any data external to the survey (for example to set priors on  IA model parameters).

The articles below are selected examples of self-calibration, ranging from its initial proposal to its application to recent survey data.

\subsubsection{Self-calibration: References}

\begin{itemize}
     \item ``A proposal on the galaxy intrinsic alignment self-calibration in weak lensing surveys"~\cite{zhang_proposal_2010}\\
    \it{First paper on IA self-calibration.}
    \item ``Self-calibration of gravitational shear-galaxy intrinsic ellipticity correlation in weak lensing surveys"~\cite{zhang_self_2010}\\
    \it{Application of self-calibration.}

    \item ``Self-calibration for three-point intrinsic alignment autocorrelations in weak lensing surveys"~\cite{troxel_self-calibration_2012}\\
    \it{Useful overview and extends the concept to three-point statistics.} 

    \item ``Self-calibration method for II and GI types of intrinsic alignments of galaxies"~\cite{yao_self-calibration_2019}\\
    \it{Review of methodology and forecast for LSST.}

    \item ``KiDS-1000: Cross-correlation with Planck cosmic microwave background lensing and intrinsic alignment removal with self-calibration"~\cite{yao_kids-1000_2023}\\
    \it{Comparison of self-calibration with other methods and application to KiDS data.}
\end{itemize} 

\section{IA Applications}\label{sec:ia_applications}

Although galaxy IA has been intensively studied as one of the most important systematic effects for \gls{weak lensing} cosmology, it can also bias \gls{RSD} measurements and has the potential to be a novel probe in cosmology.
Here we list a selection of papers that describe some of these applications.
As a direct result of galaxy formation mechanics, in principle IA can also be used to explore the relationship between the formation of galaxies and their dark matter environment.
However, currently, higher precision measurements and more detailed modeling are necessary before IA is useful for constraining galaxy formation, particularly on small scales.

\newpage
\subsection{Redshift-space distortion bias}

When IA is combined with a bias in galaxy orientations due to a survey's target selection, the survey can measure an enhanced clustering along the \gls{LOS}.
This anisotropic clustering effect mimics \gls{redshift} space distortions (\gls{RSD}), which are used to measure the growth of large-scale structure~\citep{kaiser_clustering_1987}.

\begin{itemize}
    \item ``Tidal alignments as a contaminant of redshift space"~\cite{hirata_tidal_2009}\\
    \it{First work that discussed orientation-dependent selection effects and IA on \gls{RSD} measurements.}
    
    \item ``Detection of anisotropic galaxy assembly bias in BOSS DR12"~\cite{obuljen_detection_2020}\\
    \it{Finds a correlation between galaxy orientation and anisotropic clustering in BOSS.}
    
    \item ``Impact of intrinsic alignments on clustering constraints of the growth rate"~\cite{zwetsloot_impact_2022}\\
    \it{Contains a helpful theory discussion on this effect.}
    
    \item ``Intrinsic alignment as an RSD contaminant in the DESI survey"~\cite{ lamman_intrinsic_2023}\\
    \it{Forecasts a significant impact on DESI's RSD measurements from IA.}
\end{itemize}

\subsection{Direct cosmological applications}

IA can be used to explore cosmological parameters through the impact that features such as \gls{RSD}, Baryon Acoustic Oscillations (BAO) and Primordial Non-Gaussianity (PNG) have on the IA signal.
Below is an incomplete list of papers discussing the use of IA to probe cosmology:

\begin{itemize}
\item ``Cosmological information in the intrinsic alignments of luminous red galaxies"~\cite{chisari_cosmological_2013}\\
    \it{Forecasts for the ability of ongoing \gls{spectroscopic} surveys to constrain local PNG and BAO in the cross-\gls{correlation function} of IAs and the galaxy density field.} 
    \item ``Imprint of inflation on galaxy shape correlations"~\cite{schmidt_imprint_2015}\\
    \it{The original paper on exploring PNG through IA.}

    \item ``Intrinsic alignment statistics of density and velocity fields at large scales:
    Formulation, modeling, and baryon acoustic oscillation features"~\cite{okumura_intrinsic_2019}\\
    \it{The first detection of the BAO feature on IA, in N-body simulations.}
    \item ``Imprint of anisotropic primordial non-Gaussianity on halo intrinsic alignments in simulations"~\cite{akitsu_imprint_2021}\\
    \it{Uses N-body simulations to show that IA can probe anisotropic PNG, while clustering cannot.}
    \item ``The impact of self-interacting dark matter on the intrinsic alignments of galaxies"~\cite{harvey_impact_2021}\\
    \it{Finds that self-interacting dark matter could suppress IA in a way distinct from potential baryonic effects.}

    \item ``Tightening geometric and dynamical constraints on dark energy and gravity: galaxy clustering, intrinsic alignment and kinetic Sunyaev-Zel’dovich effect"~\cite{okumura_tightening_2022}\\
    \it{Demonstration of the extra constraining power obtained by adding IA and kinetic Sunyaev-Zel’dovich to conventional clustering measurement, using Fisher matrix.}

    \item ``Constraints on anisotropic primordial non-Gaussianity from intrinsic alignments of SDSS-III BOSS galaxies"~\cite{kurita_constraints_2023}\\
    \it{First constraints on anisotropic PNG using IA.}
     \item ``The alignment of galaxies at the Baryon Acoustic Oscillation scale"~\cite{van_dompseler_alignment_2023}\\
    \it{Gives a theoretical explanation of the characteristics of the BAO feature on IA.}

\end{itemize}

\newpage

\section{Author Contributions}\label{sec:contributions}
\noindent The author order is lead, followed by the coauthors in reverse alphabetical order.
All coauthors contributed equally.\\
\textbf{C.L.}: \href{mailto:claire.lamman@cfa.harvard.edu}{claire.lamman@cfa.harvard.edu}\\
\textbf{E.T.}: \href{mailto:e.tsaprazi@imperial.ac.uk}{e.tsaprazi@imperial.ac.uk}\\
\textbf{J.S.}: \href{mailto:jingjing.shi@ipmu.jp}{jingjing.shi@ipmu.jp}\\
\textbf{N.S.}: \href{mailto:nikolina.sarcevic@gmail.com}{nikolina.sarcevic@gmail.com}\\
\textbf{S.P.}: \href{mailto:s.pyne@ucl.ac.uk} {s.pyne@ucl.ac.uk}\\
\textbf{E.L.}: \href{mailto:elegnani@ifae.es}{elegnani@ifae.es}\\
\textbf{T.F.}: \href{mailto:tassia.ferreira@physics.ox.ac.uk}{tassia.ferreira@physics.ox.ac.uk}

\section*{Acknowledgments}

\noindent Thank you to Elisa Chisari and Rachel Mandelbaum for your support and for providing invaluable feedback. We also thank the two reviewers from the Open Journal of Astrophysics for thorough suggestions.\par

\noindent Thank you to the \href{https://github.com/echo-IA}{echoIA} group and its leaders, Jonathan Blazek and Benjamin Joachimi, for creating a community which connects IA efforts across collaborations.
We also thank the organizers of the Lorentz Center workshop \textit{hol-IA: a holistic approach to galaxy intrinsic alignments} held from 13 to 17 March 2023 at Leiden University, where this project was initiated.

We thank Anya Paopiamsap for producing \rfig{cls}.

\noindent CL acknowledges support from the National Science Foundation Graduate Research Fellowship under Grant No. DGE1745303, the U.S.\ Department of Energy under grant DE-SC0013718, NASA under ROSES grant 12-EUCLID12-0004, and the Simons Foundation.
\par
\noindent ET would like to acknowledge support by the research project grant ‘Understanding the Dynamic Universe’ funded by the Knut and Alice Wallenberg Foundation under Dnr KAW 2018.0067 and the scholarship under the number PH2022-0007 by the Royal Swedish Academy of Sciences (Kungliga Vetenskapsakademien).
TF is supported by a Royal Society Newton International Fellowship (NIF-R1-221137).
NS and TF would like to acknowledge support from LSST DESC including travel for DESC infrastructure development supported by DOE funds administered by SLAC. This work was partially enabled by funding from the UCL Cosmoparticle Initiative.
EL acknowledges support from MCIN/AEI/10.13039/501100011033 and the FSE+ under the program Ayudas predoctorales of the Ministerio de Ciencia e Innovación PRE2022-101625.
JS acknowledges useful discussions with Toshiki Kurita on the 3-dimensional IA power spectrum.

\bibliography{references}


\clearpage

\printglossaries\label{sec:glossary}

\end{document}